\documentclass[fleqn,usenatbib]{mnras}

\usepackage{newtxtext,newtxmath}

\usepackage[T1]{fontenc}

\DeclareRobustCommand{\VAN}[3]{#2}
\let\VANthebibliography\thebibliography
\def\thebibliography{\DeclareRobustCommand{\VAN}[3]{##3}\VANthebibliography}

\usepackage{graphicx}	
\usepackage{amsmath}	

\usepackage{amssymb}	

\usepackage{multirow}   
\usepackage{makecell}
\usepackage{ulem}

\title[dispersion measure of FRB host galaxies]{The dispersion measure of Fast Radio Bursts host galaxies: estimation from cosmological simulations}

\author[Jian-feng Mo et al.]{
Jian-Feng Mo$^{1}$,
Weishan Zhu$^{1}$\thanks{E-mail: zhuwshan5@mail.sysu.edu.cn (WSZ)},
Yang Wang$^{2}$,
Lin Tang$^{3,1}$,
and Long-Long Feng$^{1}$
\\
$^{1}$ School of Physics and Astronomy, Sun Yat-Sen University, Zhuhai campus, No. 2, Daxue Road, Zhuhai, Guangdong, 519082, China\\
$^{2}$ Department of Mathematics and Theories, Peng Cheng Laboratory, No. 2, Xingke 1st Street, Nanshan District, Shenzhen, 518000, China\\
$^{3}$School of Physics and Astronomy, China West Normal University, No. 1, ShiDa Road, 637002, Nanchong, China}

\date{Accepted XXX. Received YYY; in original form ZZZ}

\pubyear{2015}

\begin{document}
\label{firstpage}
\pagerange{\pageref{firstpage}--\pageref{lastpage}}
\maketitle

\begin{abstract}

The dispersion measure(DM) of fast radio burst encodes important information such as its distance, properties of intervening medium. Based on simulations in the Illustris and IllustrisTNG projects, we analyze the DM of FRBs contributed by the interstellar medium and circumgalactic medium in the hosts, $\rm{DM_{host}}$. We explore two population models - tracing the star formation rate (SFR), and the stellar mass, i.e. young and old progenitors respectively. The distribution of $\rm{DM_{host}}$ shows significant differences at $z=0$ between two populations: the stellar mass model exhibits an excess at the low DM end with respect to the SFR model. The SFR (stellar mass) model has a median value of 179 (63) $\rm{pc\, cm^{-3}}$ for galaxies with $M_*=10^{8-13}\,M_{\odot}$ in the TNG100-1. Galaxies in the Illustris-1 have a much smaller $\rm{DM_{host}}$. The distributions of $\rm{DM_{host}}$ deviate from log-normal function for both models. Furthermore, two populations differ moderately in the spatial offset from host galaxy's center, in the stellar mass function of hosts. $\rm{DM_{host}}$ increases with the stellar mass of hosts when $M_*<10^{10.5}\,M_{\odot}$, and fluctuate at higher mass. At $0<z<2$, $\rm{DM_{host}}$ increases with redshift. The differences in $\rm{DM_{host}}$ between two populations declines with increasing redshift. With more localized events available in the future, statistics such as $\rm{DM_{host}}$, the offset from galaxy center and the stellar mass function of hosts will be of great helpful to ascertain the origin of FRB. Meanwhile, statistics of $\rm{DM_{host}}$ of localized FRB events could help to constrain the baryon physics models in galaxy evolution.
\end{abstract}

\begin{keywords}
radio continuum: transients -- stars: magnetars -- galaxies:haloes -- galaxies: ISM -- methods: numerical
\end{keywords}



\section{Introduction}

Fast Radio Bursts (FRB), also known as Lorimer bursts, constitute an enigmatic class of millisecond duration radio transients of extragalactic origin, which was first discovered in 2007 \citep[][]{2007Sci...318..777L}. Thanks to the efforts of multiple teams with facilities such as Parkes, ASKPA, UTMOST,CHIME, FAST etc in the past few years \citep[e.g.][]{2013Sci...341...53T, 2017MNRAS.468.3746C, 2018MNRAS.475.1427B, 2018Natur.562..386S, 2019Natur.566..230C,2021ApJ...909L...8N, 2021arXiv210604352T}, more than 500 FRB sources have been discovered so far. Tens of these sources are repeating bursts \citep[e.g.][]{2016Natur.531..202S, 2019Natur.566..230C, 2019ApJ...885L..24C, 2020ApJ...891L...6F, 2021ApJ...910L..18B}, and 19 events have been localized (\citealt{2020ApJ...903..152H} and references therein). However, the physical origin of FRB remains an open question, and a variety of theoretical models have been proposed \citep[][]{2019ARA&A..57..417C, 2019A&ARv..27....4P,2019PhR...821....1P, 2020Natur.587...45Z, 2021arXiv210710113P}. Magnetar is now believed to be the origin of some FRB events. A millisecond-duration radio burst from the Galactic magnetar SGR 1935+2154 has been detected, although its spectral energy is lower than the weakest FRB by about 40 times \citep[][]{2020Natur.587...59B, 2020Natur.587...54C}. Nevertheless, there are multiple potential pathways along which magnetar FRB sources can form \citep[e.g.][]{2019ApJ...886..110M}. In addition, models other than magnetars as the engine of FRB have not been ruled out yet.  

To unveil the origin of FRB, it is crucial to explore their physical properties including distance, luminosity, redshift distribution, host galaxies population and local environment, which can be inferred from several important features of the observed signals. The observed signals of FRB have experienced multiply propagation effects such as dispersion, scattering and polarization that caused by the intervening medium along the line of sight(L.O.S.) before reaching us \citep[][]{2019A&ARv..27....4P,2019ARA&A..57..417C}. The dispersion measure (DM) of the known FRBs has a wide distribution ranging from $\sim 100$ to 3038 $\rm{pc}\, \rm{cm}^{-3}$, which are larger than the estimated contribution from the interstellar medium in the Milky Way (MW) and the medium in MW's halo, denoted as $\rm{DM}_{\rm{MW, ISM}}$ and $\rm{DM}_{\rm{MW, Halo}}$ respectively. The extragalactic part of DM is attributed to the interstellar medium (ISM), denoted as $\rm{DM_{host,ISM}}$, and the circumgalactic medium (CGM) of host galaxies, i.e the medium outside of galaxy but within the parent halo of host galaxies, denoted as $\rm{DM_{host,halo}}$, and the intergalactic medium (IGM), denoted as $\rm{DM_{IGM}}$. The latter includes circumgalactic medium of foreground galaxies along the L.O.S.. Some work also suggest that the plasma in the local environment (a few pc) of source may make a significant contribution to the DM\citep[][]{2022Natur.606..873N,2021ApJ...922..173C}. If the contribution from the local environment of source is not included, the total DM of observed FRB usually reads as 
\begin{equation}
    \rm{DM_{obs} = DM_{MW,ISM} + DM_{MW,Halo} + DM_{IGM} + DM_{Host}/(1+z)}
\label{eqn:DM_obs}
\end{equation}
, where $\rm{DM_{host}=DM_{host,ISM}+DM_{host,halo}}$.
The MW contribution can be estimated by the NE2001 model \citep[][]{2002astro.ph..7156C} or YMW16 model \citep[][]{2017ApJ...835...29Y}. 

The knowledge of the total DM caused by the host galaxies and halo, $\rm{DM_{host}}$, are urgently needed for the following reasons. Firstly, the information of $\rm{DM_{host}}$ is very important for the estimation of the distance to FRB sources. Except for a dozen of localized events, the distances/redshifts of the other FRBs are usually inferred from the DM contributed by the IGM, using the $\rm{DM_{IGM}}-z$ relation, which have been estimated analytically, and from cosmological simulations \citep[e.g.][]{2003ApJ...598L..79I, 2004MNRAS.348..999I, 2014ApJ...780L..33M, 2015MNRAS.451.4277D, 2018ApJ...865..147Z, 2019MNRAS.484.1637J, 2020Natur.581..391M, 2021ApJ...906...95Z}. In practice, the $\rm{DM_{IGM}}$ of observed events is usually determined by subtracting $\rm{DM_{host}}$ from the extragalactic DM, i.e. $\rm{DM_{EXG}}$. The derived redshift is then further used to assess the luminosity, evolution and origin of FRB, as well as to probe cosmology. Secondly, $\rm{DM_{host}}$ are closely related to the properties of host galaxies, as well as the spatial distribution of FRB within host galaxies. A good knowledge of $\rm{DM_{host}}$ would in turn help to ascertain the origin of FRB. 

However, it is very difficult to estimate the value of $\rm{DM_{host}}$ for a given FRB event, as it requires the detailed information of the electron distribution and the position of FRB in the host galaxy. At first glance, this task seems relatively easy for those events that have been localized, since the redshifts, the properties of their host galaxies, and the position of FRB in their host galaxies are known. Yet, it is still quite challenge to estimate the electron distribution in the known host galaxies. The estimated $\rm{DM_{host}}$ of localised events ranges from $\sim 10$ to 1121 $\rm{pc \, cm^{-3}}$, but have large uncertainties \citep[e.g.][]{2017ApJ...834L...7T, 2019Natur.572..352R, 2020Natur.577..190M, 2020Natur.581..391M, 2021ApJ...910L..18B, 2021ApJ...922..173C,  2021arXiv210511445K, 2022Natur.606..873N, 2022arXiv220213458O}. For most of the FRBs, this task is more difficult as their host galaxies are even unknown. As a compromise, the statistical features of $\rm{DM_{host}}$, i.e., the distribution and typical value, are often wanted, and estimated by theoretical models. By assuming different scaled models of smooth electron distributions in galaxies, previous theoretical studies indicate that $\rm{DM_{host}}$ would follow a simple log-normal form with median value of $\sim 100\rm{pc}\, \rm{cm}^{-3}$ \citep[e.g.][]{2015RAA....15.1629X, 2018MNRAS.481.2320L, 2020A&A...638A..37W}.  This result has been adopted in many references to estimate the distance of observed FRB from DM. Alternatively, the $\rm{DM_{host}}$ is assumed to follow either a normal or a log-normal distribution, and the median value is a free parameter that ranges from 20-200 $\rm{pc\, cm^{-3}}$ \citep[e.g.][]{2020Natur.581..391M}.

Recently, tremendous success has been made in the study of galaxy formation and evolution by using cosmological hydrodynamic simulations. Several state-of-art simulation projects such as the Illustris \citep[][]{2014MNRAS.444.1518V}, EAGLE \citep[][]{2015MNRAS.446..521S, 2015MNRAS.450.1937C} and IllustrisTNG \citep[][]{2018MNRAS.475..624N}, can produce samples of galaxies that are quite similar to real galaxies in a variety of individual and statistical features. This provides an opportunity to use galaxy samples produced by simulations to estimate $\rm{DM_{host}}$.
\cite{2020AcA....70...87J} estimates the total DM contributed by the host galaxy and halo, $\rm{DM_{host}}$, using a relatively lower resolution simulation sample TNG100-3, and assuming the FRB events trace the stellar mass. \cite{2020AcA....70...87J} shows that the averaged value of $\rm{DM_{host}}$ is related to the stellar mass, being larger for more massive galaxies, and increases with the redshift. \cite{2020AcA....70...87J} also demonstrates that $\rm{DM_{host}}$ decreases with the projected distance of FRB from its host center. The value of $\rm{DM_{host}}$, averaged over varies galaxies mass and FRB positions, are $83 \, \rm{pc}\, \rm{cm}^{-3}, 198 \, \rm{pc}\, \rm{cm}^{-3}, 370$ $\rm{pc}\, \rm{cm}^{-3}$ at $z=0, 1, 2$ respectively in \cite{  2020AcA....70...87J}.

\cite{2020ApJ...900..170Z} estimated the DM caused by the medium within subhalo (galaxy) samples, i.e., $\rm{DM_{host, ISM}}$, in the TNG100-1 simulation. They selected three sets of galaxy samples within certain stellar mass and star formation rate range to mimic the host galaxies of repeating FRB 121102, and 180916, and non-repeating FRB, each with 200-1000 galaxies and 500 FRB events in each galaxy. They made an assumption of the position of FRB tracing star forming cells for repeating events, and tracing the binary neutron star mergers for non-repeating events. Their work suggested that $\rm{DM_{host, ISM}}$ can be fitted by the log-normal function, and the median value of $\rm{DM_{host, ISM}}$ ranges from $\sim 30-100\rm{pc}\, \rm{cm}^{-3}$ at z=0, and increases with increasing redshift.  \cite{2021ApJ...906...95Z} probed the $\rm{DM_{host}}$ of all the halos more massive than $10^{11} M_{\odot}$ in an adaptive mesh refined simulation, and found that the distribution of $\rm{DM_{host}}$ may deviate from the log-normal form. These works show notable discrepancy with previous theoretical models in the median value, the evolution trend with redshift, and the distribution form. Yet, the robustness of these works could be affected by various factors, such as the relatively poor resolution of the simulation sample used (e.g. \citealt{2020AcA....70...87J}), the limited galaxy samples and the assumption of a fixed FRB events for each galaxy (e.g. \citealt{2020ApJ...900..170Z}), and the assumption that mock events lie at the center of halos (e.g. \citealt{2021ApJ...906...95Z}).

Taking account of the important role of $\rm{DM_{host}}$, it is worthwhile to reexamine the estimation of $\rm{DM_{host}}$ based on cosmological hydrodynamic simulations, with large number of galaxy samples, and different models of FRB population, as well as different sets of simulations simultaneously. While magnetar is now believed to be the origin of some FRB, other progenitor models remain competitive \citep[][]{2019A&ARv..27....4P}. Also, theoretical studies suggest that there could be two primary channels to form the magnetar associated with FRB, one is from massive star progenitors directly, and the other is from relatively long time evolve and interaction of compact stars, such as merger of neutron star, and collapse of white dwarf \citep[e.g.][]{2019ApJ...886..110M, 2020Natur.587...45Z}. 

On the other hand, the distribution of gas and electron in the galaxies are highly affected by multiple baryon physics, such as gas cooling, star formation, feedback from massive star and supernova, and active galactic nuclei. Currently, these processes are implemented in simulations by various sub-grid models due to limited spatial resolution. Consequently, the
estimated $\rm{DM_{host}}$ of galaxies in simulations would depend on the adopted sub-grid models. Note that, the implemented sub-grid models for some physics, e.g. feedbacks from supernova and AGN, can vary significantly across current state-of-the-art simulations, and it is unclear which model is more close to the reality yet \citep[][]{2017ARA&A..55...59N}. For instance, starting with a very similar initial conditions, the galaxies produced by the Illustris and IllustrisTNG projects display notable differences in galaxies sizes, gas properties \citep[][]{2018MNRAS.473.4077P}. Thus, it is necessary to investigate the impact of the difference in sub-grid models across different simulations on the estimated $\rm{DM_{host}}$. Recently, \cite{2021arXiv210608618K} use zoom-in simulations of Milky Way-mass galaxies to probe the effect of two different galaxy formation models - EAGLE and Auriga, and find that different sub-grid models can lead to different gas properties, and hence different dispersion measures. In turn, more reliable statistics of $\rm{DM_{host}}$ based on future observation may help to constrain the sub-grid baryon physical models.


In this work, we use galaxy samples from both the Illustris and IllustrisTNG simulation projects to estimate $\rm{DM_{host}}$, and investigate the impact of systematic effects such as FRB population models, sub-grid physics models in different simulations, the assumed number of events in different galaxy. We also study the dependency of $\rm{DM_{host}}$ on the stellar mass, morphology of host, and the redshift. This paper is laid out as follows. In Section 2, we briefly introduce the simulations and galaxy samples, and describe our methods used for placing the mock FRB events in galaxies, and calculating the dispersion measure contributed by host galaxies and halos. Section 3 presents the statistical analysis. We discuss our results in Section 4. Finally, we summary our findings and make concluding remarks in Section 5.

\section{Methods}
\label{sec:method}


\subsection{Simulations}
\label{sec:maths} 

We carry out our investigation based on the public released data of the Illustris-1 (hereinafter referred to as Illustris) simulation from the Illustris project \citep[][]{2014MNRAS.444.1518V}, and the TNG100-1 (hereinafter referred to as TNG) simulation from the IllustrisTNG project \citep[][]{2018MNRAS.475..624N}. Both simulations are run by the moving mesh code AREPO \citep[][]{2010MNRAS.401..791S}. The initial conditions for these two simulations are almost the same, except the moderate differences in the cosmological parameters. Illustris adopts the WMAP9 cosmology with $\Omega_m =\Omega_{dm} + \Omega_b = 0.2726$, $\Omega_{\Lambda}$ = 0.7274, $\Omega_b$ = 0.0456, $\sigma_8$ = 0.809, $n_s$ = 0.963, and $H_0$ = 100$h \ \rm{km}\ {s}^{-1} \rm{Mpc}^{-1}$ with $h$ = 0.704 \citep[][]{2013ApJS..208...19H}, while TNG use the Planck constraints on cosmology parameters, i.e., $\Omega_m = 0.3089$, $\Omega_b$ = 0.0486, $\Omega_\lambda$ = 0.6911, $\sigma_8$ = 0.8159, $n_s$ = 0.9667, $h$ = 0.6774 \citep[][]{2016A&A...594A..13P}. These two simulations adopt the same cubic box size with a side length of $75/h$ cMpc. The mass resolutions of dark matter particles are $7.5 \times 10^6 M_{\odot}$ and $6.3 \times 10^6 M_{\odot}$ in Illustris and TNG respectively. For baryonic matter, the mass resolutions are $1.4 \times 10^6 M_{\odot}$ and $1.3 \times 10^6 M_{\odot}$ in Illustris and TNG respectively.  

The sub-grid physics implemented in the TNG and Illustris simulations are different mainly in the following aspects, the galactic winds driven by stellar feedback \citep[][]{2018MNRAS.473.4077P}, and the evolution and feedback of super-massive black holes \citep[][]{2017MNRAS.465.3291W}. The revised models in TNG can suppress the star formation more effectively, and lead to a better agreement with the observations than Illustris on the cosmic star formation rate after $z=1$, on the stellar content and sizes of galaxies at the low mass end, and on the gas fraction in massive dark matter halos. We refer the readers to \cite{2017MNRAS.465.3291W} and \cite{2018MNRAS.473.4077P} for more details about the differences in feedback models between the two simulations. 

\begin{table*}
	\centering
	\caption{The number of selected galaxy samples and the total number of mock FRBs events at $z=0$ }
	\label{tab:num galaxies and FRBs and in disk}
	\begin{tabular}{ccccccc} 
		\hline
		simulation & model & \makecell[c]{galaxies \\ number}  & \makecell[c]{total FRB \\ number} & \makecell[c]{Disk \\ number}  & \makecell[c]{total FRB \\ number in disk}  & \makecell[c]{ number of L.O.S. \\ for each FRB}  \\
		\hline
		\multirow{3}{5em}{TNG100-1}  & PSFR-VNum & 33617 & 404444  & 11278 & 302061 & 20\\

		                       & PStar-VNum   & 37827 & 294840  & 12496 & 188746 & 20\\

		\hline
		\multirow{3}{5em}{Illustris-1} & PSFR-VNum   & 63692 & 686697  & 6364 & 308999 & 20\\
		                        
		                        & PStar-VNum  & 70049  & 380599  & 6704 & 162814  & 20\\

		\hline
	\end{tabular}
\end{table*}

\subsection{Galaxy samples}
\label{subsec:samples}
Up to now, 19 FRB have been localized (e.g. \citealt{2020ApJ...903..152H, 2021ApJ...919L..23F, 2021arXiv210801282B, 2021ApJ...910L..18B, 2021ApJ...919L..24B}). The details of these FRB-host galaxies can be found in the online database \footnote{\href{https://frbhosts.org/}{https://frbhosts.org/}}. These galaxies cover a broad, continuous range of stellar mass ($10^8 - 8 \times 10^{10} M_{\odot}$), star formation rate (0.03-10 $M_{\odot} yr^{-1} $), and color. Considering the diversity of known host galaxies, subhalos with stellar mass $M_*$ above $10^8 M_{\odot}$ and gas mass $M_{gas}>0$ in Illustris and TNG are included in the host galaxy samples in our study. Namely, our study focus on the $\rm{DM_{host}}$ of galaxies with $M_*>10^8 M_{\odot}$. The total number of host galaxies at $z=0$ in two simulations are listed in Table \ref{tab:num galaxies and FRBs and in disk}.
In fact, we cannot rule out the probability that FRB can happen in a galaxy with stellar mass less than $10^8 M_{\odot}$, or happen in some place out side of dark matter halos. We find that around $\sim 8.5\%$ (7.5 $\%$) of the SFR (stellar mass) of the whole simulation volume are outside of our galaxies samples. These SFR regions and stellar particles are either hosted by mini halos or out side of halos, FRBs associated to these regions (stellar particles) would have a median $\rm{DM_{host}}$ smaller than our samples. We would like to remind the readers that our results of $\rm{DM_{host}}$ are based on galaxies samples with $M_*>10^8 M_{\odot}$. The median value of $\rm{DM_{host}}$ is expected to decrease moderately, if all the SFR and stellar mass in the entire simulation volume are included. 

On the other hand, the Subfind algorithm used by the Illustris and IllustrisTNG projects to identify subhalos cannot discriminate non-cosmological objects, such as fragments or clumps, which can be produced by internal processes (e.g. disk instabilities), from galaxies with cosmological origin. The IllustrisTNG project offers a new scheme to tackle this issue. We adopt the criteria of `SubhaloFlag == True' in the TNG100 data when selecting galaxies from the subhalo catalogue. However, this criteria can't be applied to subhalos in the Illustris project. If we don't apply this criteria, the number of selected galaxy samples in TNG100-1 would be overestimated by $\sim 6\%$. The corresponding fraction should be close in the Illustris simulation. We have inspected the impact of `fake' galaxies, i.e. fragments or clumps, on the distribution and median value of $\rm{DM_{host}}$ for the TNG samples. We find that including these `fake' galaxies would slightly change the distribution of $\rm{DM_{host}}$ and increase the median value by $\sim 1\% - 3\%$. For the sake of conciseness, figures are not presented. Therefore, the contamination due to fake galaxies in Illustris should have a limited effect on our results.

\subsection{Population of Mock Fast Radio Bursts}
\label{sub:FRB population}

For some of the proposed models of FRB progenitor, such as magnetar originated from collapse of massive star, FRB sources would trace the star formation activity in the host galaxy. Meanwhile, for some other progenitor models including magnetar formed by merger of neutron stars or collapse of white dwarfs, FRB would trace the stellar mass. Correspondingly, we use two models of mock FRB population in our study, assuming that the mock FRB events are related to the stellar mass, and star formation rate respectively in galaxy samples compiled from simulations. These two population models will be denoted as `PStar' and `PSFR' population models respectively in the following context. For the latter population model, we only use mock galaxies with a specific star formation rate (sSFR), i.e., $\rm{SFR}/M_*$, above $10^{-11} yr^{-1}$, below which the star formation activity are usually considered as quenched. Thus, the number of host galaxy samples with `PSFR' population model is smaller by $\sim 10\%$. 

The method that we place mock samples of fast radio bursts into galaxies are as follows. Firstly, we determine the number of mock FRB in each galaxy. The statistical results of $\rm{DM_{host}}$ may affected by the number of FRB placed in each galaxy, either if the number is insufficient or if it is not well sampled. In the literature, some previous study assume a constant number of mock FRB events in each galaxy \citep[][]{2020ApJ...900..170Z}, which is somehow oversimplified. To probe the effect of this factor, we adopt two different scenarios to set the number of FRB in each galaxy. In the first scenario, the number of mock events are scaled to the stellar mass of host galaxies for the 'PStar' population model, and scaled to the star formation rate of hosts for the 'PSFR' population model respectively. This scenario is denoted as 'VNum', and is combined with the name of population model in the following context as 'Pxxx-VNum' to identify various models. More specifically, for the `PStar-Vnum' model, the number of mock FRB in a given host galaxy with stellar mass $M_*$ is decided by 
\begin{equation}
     N_{\rm{FRB}} = min[int \left(\frac{M_*} {10^9 M_{\odot}} \right)+1, \  100]
     \label{eqn:number FRB of Mstar}
\end{equation}
. For the `PSFR-VNum' model,  
the number of mock FRB in a given host galaxy with star formation rate $\Dot{M_*}$ is set to   
\begin{equation}
    N_{\rm{FRB}} = min[int \left(\frac{\Dot{M_*}}{0.033 M_{\odot}/yr} \right)+1, \  100]
    \label{eqn:number FRB of SFR}
\end{equation}
In practice, we use an upper limit of 100 events in one galaxy to save the computation time. Because only 2\% of galaxies have 100 FRB mock events according to eqn. \ref{eqn:number FRB of Mstar} and \ref{eqn:number FRB of SFR}, the usage of this upper limit should have minor effect on our results. The requirement that all selected galaxy samples have at least one FRB seems to lead to a bias for small/low SFR galaxies. However, this bias is weak because small/low SFR galaxies account a small fraction of the total stellar mass/SFR in the simulation volume. We have evaluated this factor by excluding those galaxies with stellar mass ranging from $10^8M_\odot$ to $10^9M_{\odot}$ for the PStar-VNum model, and those with SFR ranging from 0 to $0.033M_\odot/yr$ for the PSFR-VNum model. We find that the median $\rm{DM_{host}}$ will increase $2\%-3\%$, with respect to the one including these small/low SFR galaxies.

The reasons that we choose the values of denominator terms in eqn. \ref{eqn:number FRB of Mstar} and \ref{eqn:number FRB of SFR} are as follows. One is to produce a sufficient number of mock events to make the statistical results reliable, while keep the required computation resource affordable. The other is to make the total mock FRB events in two population models to be similar, and close to the number in \cite{2020ApJ...900..170Z}. The total numbers of mock FRB events placed in the galaxy samples we selected from the Illustris and TNG simulation with the `PStar-VNum' and `PSFR-VNum' models range from nearly 300,000 to 680,000, and are listed in table \ref{tab:num galaxies and FRBs and in disk}. Note that, a more realistic sampling should consider estimation on the event rate of FRB events via different formation channel. For instance, the rate of binary neutron star merger, which is an alternative channel for FRB events, have been estimated in the literature based on simulation and observation. We discuss this in section \ref{subsec:spatial coincident}.

In the second scenario, we set the number of FRB in each galaxy to a constant, regardless of the FRB population models. This scenario is named as 'CNum', and also will be combined with the name of population models for identification as `PStar-CNum' and `PSFR-CNum'. In these two models, a fixed number of 12 mock FRB is assigned to each galaxy to make the total number of FRB approximately equals to the `VNum' scenario at redshift z=0. The main context will focus on the results of `VNum' models, while the results of `CNum' are shown and discussed in the Appendix.

After the number of mock FRB in each galaxy is determined, we then set up the position of each FRB in each galaxy. For the `PStar' population model, the spatial distribution of FRB is assumed to follow the stellar mass distribution. Namely, the probability of FRB locating in a particular cell is proportional to the stellar mass in this cell. While for the `PSFR' population model, the probability of FRB occurring in a given cell is proportional to its local star formation rate. More specifically, the probability of a FRB locating in the $i$th cell, $P_i$, is given by 
\begin{equation}
    P_i = \frac{M_{*,i}}{\sum M_{*,i}}, 
\end{equation}
and 
\begin{equation}
 P_i = \frac{SFR_{i}}{\sum SFR_{i}}
\end{equation} for the PStar and PSFR model,
where $M_{*,i}, \, SFR_i$ are the stellar mass, star formation rate of $i$th cell respectively. Every mock event is placed at the center of the cell it locating in.

\subsection{calculation of dispersion measure}
\label{sub:dm_calc}

The dispersion measure can be written as
\begin{equation}
    \rm{DM}=\int{n_e}\it{dl}
\label{eqn:dm}
\end{equation}
, where $n_e$ is the number density of electron along the line-of-sight and $dl$ is the integral element of line-of-sight. 

In this work, we probe the DM contributed both by the ISM of host galaxy, i.e. electron in subhalo, and by the CGM in host halo, i.e. electron outside of subhalo but within the parent halo. Usually, the FRB events are offset from the center of their host galaxies. In addition, both the ISM in galaxy and CGM in halo are usually non-homogeneous. Consequently, there will be significant fluctuation on the estimated $\rm{DM_{host}}$ among different line-of-sights. Taking into account these factors, we draw 20 line-of-sights along random directions for each mock FRB source. At $z=0$, the total number of L.O.S. produced for the models ranges from 6 to 14 millions, which are listed in Table \ref{tab:num galaxies and FRBs and in disk}. Each line-of-sight starts from the place where a mock FRB sites in, i.e., the center of the cells they are placed, and ends at the boundary of the parent dark matter halo, which is the sphere centered at halo center and within which the density is 200 times of the mean density of the Universe. The radius of halo is denoted as $\rm{R_{m200}}$. The shape of subhalo is usually irregular. For the sake of simplicity, we assign the gas and electron within the sphere of a radius five times of the half-mass radius of stars ($\rm{R}_{50}$) as the ISM in the host galaxy, and those gas and electron within the shell between $5 \times \rm{R}_{50}$ and $\rm{R_{m200}}$ as the CGM in the host halo. That is, we integrate eqn. \ref{eqn:dm} over each line-of-sight to evaluate the DM from the ISM in the host galaxy, i.e $\rm{DM_{host, ISM}}$, and the CGM in the host halo, i.e. $\rm{DM_{host, halo}}$.  
The reason why we use $5 \rm{R}_{50}$ as the boundary between ISM and CGM  is to make the results based on the samples in TNG to be comparable to observational and theoretical studies on $\rm{DM_{host,ISM}}$ of the Milky Way. The median $\rm{R}_{50}$ of all the galaxy samples as mentioned above are approximately 2.7 kpc, 6.5 kpc in TNG and Illustris respectively. For galaxies with stellar mass similar to Milky Way, the median $\rm{R}_{50}$ are 4.8 kpc, 6.2 kpc for PStar, PSFR host galaxies respectively in TNG simulation. The corresponding sizes are 8.8 kpc, 9.2 kpc respectively for Illustris samples.
In comparison, observational and theoretical works in the literature usually account electron within the radius of $\sim 20$ kpc when estimating $\rm{DM_{host,ISM}}$ of the Milky Way (e.g. \citealt{2019MNRAS.485..648P, 2020ApJ...888..105Y}).

In the Illustris and TNG simulations, an unstructured Voronoi tessellation was used for adaptive spatial discretization. Information of each Voronoi cell at different redshifts can be obtained from simulation snapshots. The electron number density is assumed to be uniform in each Voronoi cell. In practice, we cut the rays from the mock events to halo boundary into pieces of segment with uniform length $\rm{\Delta L}$, and calculate the integration in eqn. \ref{eqn:dm} along a given line-of-sight as
\begin{equation}
    \rm{DM} =\sum{\Delta \rm{DM_i}} \simeq \sum{n_{e,i} \cdot \Delta L}
\label{eqn:dm_diff}
\end{equation}
, where $\Delta \rm{DM_i}$ is the contribution from $i$th segment, $\rm{n_{e,i}}$ is the electron density at the center of the $i$th segment and is approximately set to the electron density at the nearest mesh-generating point (i.e. the 'particle/cell' in simulation). In order to balance accuracy and computational efficiency, we set $\rm{\Delta L = 1\, kpc/h}$ for all the line-of-sights. The result DM along a particular L.O.S is divided into two parts. 
The sum of these two components is denoted as $\rm{DM_{host}}$, which provides a more reasonable estimation on the DM of FRB that caused by the host, and is used to compare with observations.

The electron number density in a Voronoic cell is derived as follows. For non-star-forming cells,
\begin{equation}
    n_e = \eta_e \cdot n_H = \eta_e X_H \frac{\rho}{m_H}
\label{eqn:ne}
\end{equation}
, where $\eta_e$ is the abundance of electron, $n_H$ is the number density of hydrogen, $X_H$ is the abundance of hydrogen, $\rho$ is the mass density in gas cell, $m_H$ is the mass of hydrogen atom. Note that, the ionization of Helium is not taken into account in eqn. ~\ref{eqn:ne}. The estimated DM should be multiplied by a factor of 1.083, if the Helium is ionised. Following previous study, we skip the contribution from the ionisation of Helium.
For star-forming cells, a sub-grid treatment of star formation and the interstellar medium was implemented in both the Illustris and TNG projects. Hence, we follow the method in \cite{2019MNRAS.483.5334S} to calculate electron number density of star-forming cells\footnote{\href{https://github.com/arhstevens/Dirty-AstroPy}{https://github.com/arhstevens/Dirty-AstroPy}}. More specifically, it divides the star-forming cells into a hot gas component (fully ionized) and a cold gas cloud component (neutral). Only the hot gas component contributes to the electron number density. According to internal energy and density of star-forming cells, we can calculate the fraction of neutral gas, i.e., $f_{neu}$. Then the electron number density in star-forming cells is given by
\begin{equation}
    n_e  = \eta_e X_H \frac{\rho}{m_H} \cdot \left(1-f_{neu} \right)
\label{eqn:ne_tot}
\end{equation}
where $f_{neu} = 0$ for non-star-forming cells.

\subsection{Convergence of DM calculation}
\label{sub:DM cal tests}

There are various factors that might affect the accuracy and overall distribution of the calculated $\rm{DM_{host}}$, such as the assumed number of mock FRB in each galaxy, the number of sightlines drawn for each mock FRB event, the total number of mock FRB events, as well as the choose of segment length when calculating the integration of DM along sightlines. We have run some tests to inspect the convergence of our results, and determine the corresponding parameter values in our calculating procedures. 

Firstly, we have investigated the impact of different total number of mock FRB events associated with galaxies in the TNG and Illustris simulations, which differs by $\sim 20\%-50\%$ as shown in Table \ref{tab:num galaxies and FRBs and in disk}. We randomly select a number of light-of-sights (L.O.S.) in the Illustris samples that equals to the number of L.O.S. in the TNG sample, and then probe the distribution of $\rm{DM_{host}}$. We find that this change has minor effect on the distribution of $\rm{DM_{host}}$. As section \ref{sub:FRB population} has introduced, we also have studied the effect of the assumed FRB number in each galaxy. In our `VNum' population models, the number of FRB events in a galaxy is related to its stellar mass or SFR. However, \cite{2020ApJ...900..170Z} adopt a constant number of 500 FRB events in each host galaxy. As a comparison test, our `CNum' model assign 12 FRB to each galaxy. The results of `CNum' on $\rm{DM_{host}}$ can be find in the Appendix. In short, we find the `VNum' model will increase the median $\rm{DM_{host}}$ by a factor of 1.4 to 2.4, with respect to the `CNum' model, as Figure \ref{fig:test of CNum model} shows.

Besides, we have probed the impact of different definition of halo boundary, $R_h=\rm{2R_{m200}}$ against $\rm{R_{m200}}$, on the estimated $\rm{DM_{host}}$. We find that the median value of $\rm{DM_{host,ISM}}$ is not changed, but the median value of $\rm{DM_{host,halo}}$ with $R_h=\rm{2R_{m200}}$ is slightly larger than that with $R_h=\rm{R_{m200}}$, which leads to the median $\rm{DM_{host}}$ of the former case is slightly larger than that of the latter, as Figure \ref{fig:test of 2RM200} shows. On the other hand, the probability distribution function at low DM end for the case with $R_h=\rm{2R_{m200}}$ shows a tail that is shallower than that with $R_h=\rm{R_{m200}}$.

Furthermore, we have studied the choice of the segment length along line-of-sights, $\Delta L$, that used in our calculation of $\rm{DM_{host}}$, i.e., eqn. \ref{eqn:dm_diff}. The gas particles/cells in simulations have different size. The average size of gas/star forming cells in TNG100-1 are $\bar{r}_{cell}=15.8 \, \rm{kpc}$, and $\bar{r}_{cell,SF}=0.355 \, \rm{kpc}$ respectively. The minimum cell size can reach 14 $\rm{pc}$. The corresponding values in Illustris-1 are similar. In our procedure, the electron density at each segment with length $\Delta L$ along the rays is approximately set to the corresponding value of the nearest cell. Therefore, a $\Delta L$ in eqn. \ref{eqn:dm_diff} is large enough such that the estimated $\Delta \rm{DM}$ of some piece of segment will be inaccurate, and hence reduce the accuracy of estimated $\rm{DM_{host}}$. Meanwhile, a small value of $\Delta L$ would consume too much computational resources. We have tested our DM calculation with different values of $\Delta L=0.1,\, 0.5,\, 1,\, 2 \, \rm{kpc/h}$. We find that $\Delta L=1 \, \rm{kpc/h}$ can balance the accuracy and computation resources consumed. In addition, We also have tested the number of sightlines drawn for each mock FRB event. The ISM and CGM in galaxies and halos usually are inhomogeneous and anisotropic. If the number of L.O.S draw for each event is too small, it might under-sampling the distribution of electronics in the host galaxy and halo.  We vary the number of L.O.S draw for each mock FRB event, and find that 20 random L.O.S. is enough to obtain a robust estimation of $\rm{DM_{host}}$ for our samples. 

Finally, we have explored the impact of simulation resolution and volume. We use the same procedures to evaluate the $\rm{DM_{host}}$ of galaxies samples in the TNG100-3 and TNG300-1 simulation. Among these three simulations, the resolution in TNG100-1 (TNG100-3) is the highest (lowest). For galaxies within the stellar mass range $10^9-10^{11} \rm{M_{\odot}}$, TNG100-3 and TNG300-1 samples have a median $\rm{DM_{host}}$ larger than the TNG100-1 samples by $\sim 40 \%$ and $\sim 10 \%$ respectively with the PSFR-VNum model. The opposite trend is observed for the PStar-VNum model, and the difference is $\sim 20\%$ between TNG100-1 and the other two simulations. Calculation based on a simulation with poor resolution will overestimate (underestimate) $\rm{DM_{host}}$ of PSFR-VNum (PStar-VNum) model. The details of the results of the tests mentioned in this chapter can be found in Table \ref{tab:fit param with DMtests} in the Appendix.

\begin{figure*}
	\includegraphics[width=1.9\columnwidth]{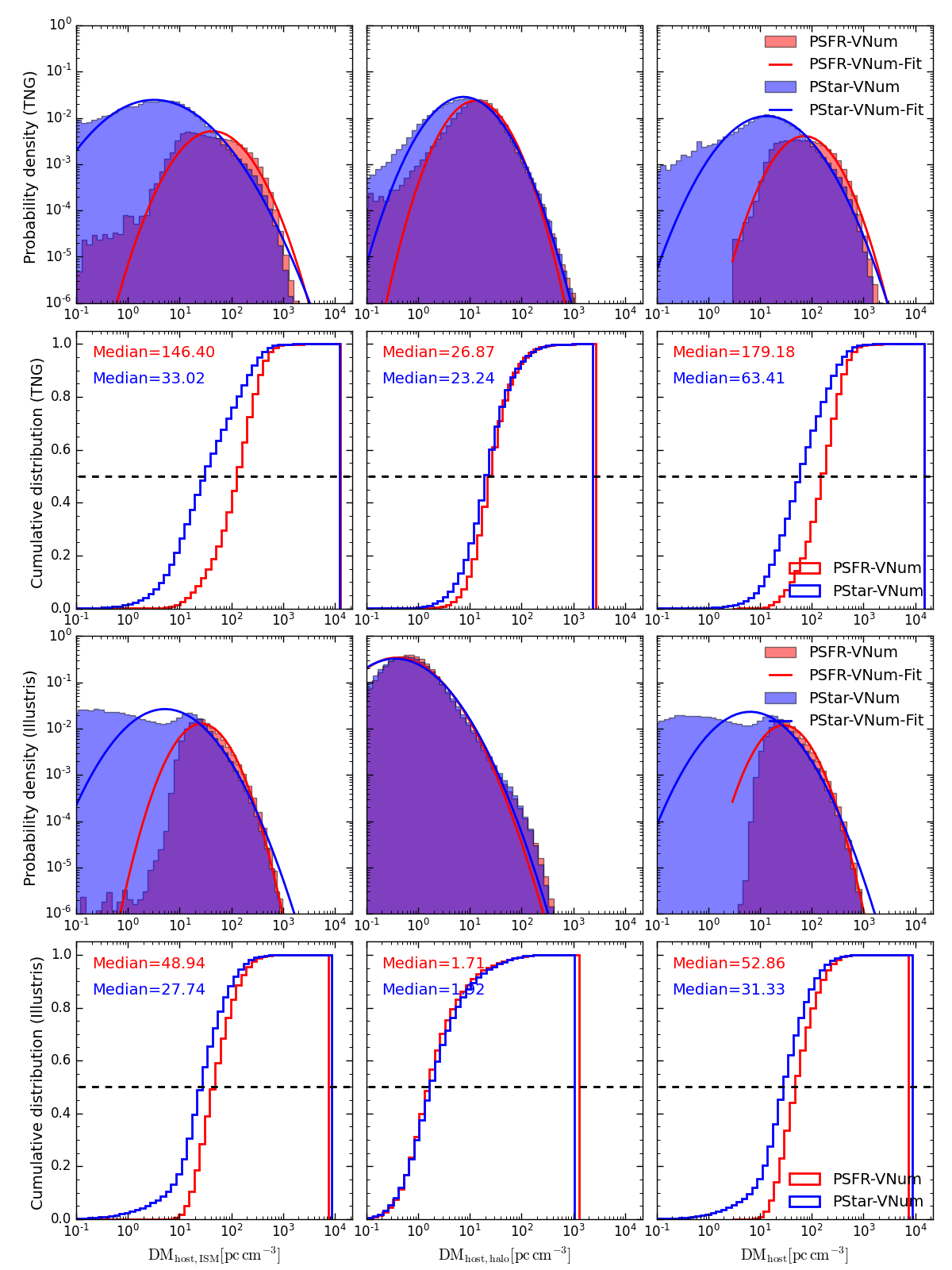}
    \caption{The probability distribution and cumulative distribution of $\rm{DM_{host}}$ of galaxies samples in two simulations at $z=0.0$. The left, middle and right columns are contribution from the ISM in host galaxies, $\rm{DM_{host, ISM}}$, from CGM in the parent halo, $\rm{DM_{host, halo}}$, and their sum $\rm{DM_{host}}$ respectively. Top (bottom) two rows: the results based on TNG100-1 (Illustris-1). Red and blue color indicate 'PSFR-VNum' and 'PStar-VNum' models (see section \ref{sub:FRB population} for detail)  respectively. Their median values are shown by the red and blue numbers. Red (blue) solid lines are the fitting results by using equation \ref{eqn:log-normal} for 'PSFR-VNum' ('PStar-VNum') model. The best fitting parameters and median value of DM distribution are shown in Table \ref{tab:fit param mainly}. }
    \label{fig:compare model in two simu}
\end{figure*}

\begin{table*}
	\centering
	\caption{The median value of $\rm{DM_{host}}$ and its two components $\rm{DM_{host, ISM}}$ and $\rm{DM_{host, halo}}$, derived from galaxies samples in the TNG100-1 and Illustris-1 simulations by different models. $e^{\mu}$ and $\sigma$ are the best fitting parameters with log-normal function for the distribution of $\rm{DM_{host, halo}}$. The last column shows the mean p-value of 20 times Shapiro–Wilk normality test for $ln(\rm{DM})$ with 1000 randomly selected mock events.}
	\label{tab:fit param mainly}
	\begin{tabular}{m{5em} m{7em} m{4em} m{4em} m{5em} m{4em} m{3em}} 
		\hline
    		simulation & model & component & median [$\rm{pc\,cm^{-3}}$] & $ e^{\mu}$[$\rm{pc\,cm^{-3}}$] &       $  \sigma  $  & p-value \\
		\hline
		\multirow{6}{5em}{TNG100-1} & \multirow{3}{7em}{PSFR-VNum} & $\rm{DM_{host,ISM}}$  & 146.40 & 126.95  & 1.03 & 0.00 \\
		                                  &           & $\rm{DM_{host,halo}}$  & 26.87 & 28.22  & 0.89 & 0.00 \\
		                                  &           & $\rm{DM_{host}}$  & 179.18 & 163.23 & 0.91 & 0.00 \\
		                        
		                        & \multirow{3}{7em}{PStar-VNum} & $\rm{DM_{host,ISM}}$  & 33.02 & 34.54 & 1.53 & 0.00 \\
		                                              &    & $\rm{DM_{host,halo}}$ & 23.24 & 23.13 & 1.06  & 0.11 \\
		                                              &    & $\rm{DM_{host}}$ & 63.41 & 63.55 & 1.25  & 0.00 \\
		\hline
		\multirow{6}{5em}{Illustris-1} & \multirow{3}{7em}{PSFR-VNum} & $\rm{DM_{host,ISM}}$ & 48.94 & 52.62  & 0.83 & 0.00 \\
		                                            &           & $\rm{DM_{host,halo}}$ & 1.71 & 2.06 & 1.30 & 0.00 \\
		                                            &           & $\rm{DM_{host}}$ & 52.86 & 56.90  & 0.83  & 0.00 \\
		                        
		                        & \multirow{3}{7em}{PStar-VNum} & $\rm{DM_{host,ISM}}$   & 27.74 & 26.52 & 1.28  & 0.00 \\
		                                              &    & $\rm{DM_{host,halo}}$ & 1.92 & 2.25 & 1.35   & 0.00 \\
		                                              &    & $\rm{DM_{host}}$ & 31.32 & 29.85 & 1.24  & 0.00 \\ 
		
		\hline
	\end{tabular}
\end{table*}

\section{dispersion measure of host galaxies and halos}
\label{sec:DMs main results}
\subsection{results at $z=0.0$}
Figure \ref{fig:compare model in two simu} shows the probability distribution function (PDF) and cumulative distribution function(CDF) of $\rm{DM_{host, ISM}}$, $\rm{DM_{host, halo}}$ and $\rm{DM_{host}}$ in the 'PSFR-VNum' and 'PStar-VNum' models, based on galaxy samples in the TNG100-1 and Illustris-1 simulations at redshift z=0. 

Figure \ref{fig:compare model in two simu} indicates that there are significant differences between the 'PSFR-VNum' and 'PStar-VNum' models in the DM contributed by host galaxies and halos. For each population model, there are also notable differences between the PDF and CDF of $\rm{DM_{host, ISM}}$ based on galaxies selected from the two different simulations. We will investigate the probable reasons that lead to those differences in the next subsection. The values of $\rm{DM_{host, ISM}}$, $\rm{DM_{host, halo}}$ and $\rm{DM_{host}}$ with the 'PStar-VNum' model mainly range from $10^{-2}$ to $10^{3}$ $\rm{pc \ cm^{-3}}$, with some rare cases of $\rm{DM_{host}}$ can be as large as $\sim 2\times 10^{4} \, \rm{pc\, cm^{-3}}$.  While for the 'PSFR-VNum' model, $\rm{DM_{host, ISM}}$ ranges from $10^{-1}$ to $10^{3}$ $\, \rm{pc \ cm^{-3}}$, $\rm{DM_{host, halo}}$ and $\rm{DM_{host}}$ ranges from $10^{0}$ to $10^{3}$ $\,\rm{pc \ cm^{-3}}$. A considerable fraction of the mock events in the `PStar-VNum' model have $\rm{DM_{host} <10.0 \,pc \ cm^{-3}}$. Such cases are rare in the 'PSFR-VNum' model. At the high DM end, the probability distribution function of $\rm{DM_{host}}$ is very similar between the two population models.

The predicted ranges of $\rm{DM_{host}}$ with either the `PStar-VNum' or `PSFR-VNum' model are generally consistent with \cite{2020AcA....70...87J}, \cite{2020ApJ...900..170Z} and \cite{2021ApJ...906...95Z}, and can cover the estimated $\rm{DM_{host}}$ of localized FRB, which ranges from as small as $10-25\, \rm{pc \, cm^{-3}}$ for event FRB 20181030A \citep[][]{2021ApJ...919L..24B}  and FRB 20200120 \citep[e.g.][]{2021ApJ...910L..18B, 2021arXiv210511445K}, to as large as $903-1121 \rm{pc\,cm^{-3}}$ for event FRB 190520B \citep[][]{2022Natur.606..873N, 2022arXiv220213458O}. Note that, there are considerable uncertainties on the estimated $\rm{DM_{host}}$ of FRB 20181030A with $\rm{DM_{obs}=103.5} \rm{pc \ cm^{-3}}$, and FRB20200120 with $\rm{DM_{obs}=87.8} \rm{pc \ cm^{-3}}$. It's possible that the $\rm{DM_{host}}$ of their hosts can be smaller than 10 $\rm{pc \ cm^{-3}}$. We will revisit this point in the discussion section. In short, current observations might miss out some FRB events with $\rm{DM_{host} > 1000\ pc \ cm^{-3}}$, and with $\rm{DM_{host} < 10\ pc \ cm^{-3}}$.

The median value of $\rm{DM_{host,ISM}}$, $\rm{DM_{host,halo}}$ and $\rm{DM_{host}}$ in the different models are illustrated by the legends in Figure \ref{fig:compare model in two simu}, and are also listed in Table \ref{tab:fit param mainly}. For galaxies samples in the TNG100-1 simulation, the median $\rm{DM_{host,ISM}}$, $\rm{DM_{host,halo}}$ and $\rm{DM_{host}}$ with the 'PSFR-VNum' model are 146.40, 26.87 and 179.18 $\rm{pc \ cm^{-3}}$ respectively. The value of $\rm{DM_{host,ISM}}$ derived here is mildly larger than the estimated result 110.96 $\rm{pc \ cm^{-3}}$ of FRB 180916 like event in \cite{2020ApJ...900..170Z}. The selected 200 subhalos for FRB 180916 in \cite{2020ApJ...900..170Z} have stellar mass $10^9-2\times 10^{11} \, M_{\odot}$, and SFR $0-2 \, M_{\odot}/yr$, which are moderately different from our samples with $M_*>10^8 \, M_{\odot}$, and sSFR $>10^{-11}\, M_{\odot}/yr$.  As we will see later, $\rm{DM_{host}}$ depends on the stellar mass, which would contribute to the differences between our results and \cite{2020ApJ...900..170Z}. In addition, a constant number of mock events in each galaxy is adopted in \cite{2020ApJ...900..170Z}, which also contributes partially to the differences. 
The median values of $\rm{DM_{host,ISM}}$, $\rm{DM_{host,halo}}$ and $\rm{DM_{host}}$ with the 'PStar-VNum' model are relatively small, i.e., 33.02, 23.24 and 63.41 $\rm{pc \ cm^{-3}}$ respectively. Meanwhile, the mean value of $\rm{DM_{host}}$ for our 'PStar-VNum' model is 127 $\rm{pc \ cm^{-3}}$, which is larger than the averaged value of $83 \pm 53\, \rm{pc \ cm^{-3}}$ at $z=0$ for TNG100-3 galaxies samples with $10^{7.5}-10^{11} \, M_{\odot}$ in \cite{2020AcA....70...87J}. To make a more direct comparison, we have applied our calculation to galaxies with stellar mass range $10^{7.5}-10^{11} \, M_{\odot}$ in TNG100-3, and the mean and median value of $\rm{DM_{host}}$ at z=0 are 116 and 51 $\rm{pc\,cm^{-3}}$ repectively. The discrepancy between our results and \cite{2020AcA....70...87J} should be caused by the differences in the calculating procedures.

For galaxies in the Illustris-1 simulation, $\rm{DM_{host}}$ obtained from the 'PStar-VNum' model is also significantly smaller than that from the 'PSFR-VNum' model. Yet, the estimated $\rm{DM_{host}}$ of galaxies in Illustris-1 are much smaller than that in TNG100-1, which holds true for both the two FRB population models considered here. For galaxies in Illustris-1, the median value of $\rm{DM_{host, ISM}}$, $\rm{DM_{host,halo}}$ and $\rm{DM_{host}}$ are 48.94, 1.71 and 52.86 $\rm{pc \ cm^{-3}}$ respectively when the FRB events trace the SFR. It will be 27.74, 1.92 and 31.33 $\rm{pc \ cm^{-3}}$ respectively when the FRB events trace the stellar mass. 

In the literature, the DM of FRBs contributed by the host is often assumed to follow a log-normal distribution. We try to fit the PDF of DM for different population models by the log-normal function, i.e.
\begin{equation}
    P(x; \mu, \sigma) = \frac{1}{x\sigma \sqrt{2\pi}} \rm{exp}\left( -\frac{(\rm{ln} \it(x)-\mu)^2}{2\sigma^2} \right)
\label{eqn:log-normal}
\end{equation}
where $\mu$ and $\sigma$ are the fitting parameters, and the mean and variance value are $ e^{\mu}$ and $ \left( e^{\sigma ^2} -1 \right)e^{\left(2\mu + \sigma^2 \right)}$ respectively. The python package 'scipy.stats.lognorm.fit' is used to do the fitting, whose estimation method is Maximum Likelihood Estimation (MLE).

We firstly ran the fitting in the range $\rm{DM_{host}} > 0$. The solid lines in Figure \ref{fig:compare model in two simu} indicate the best fitting results. 
The fitting parameters are listed in Table \ref{tab:fit param mainly}. 
Since it is quite difficult to obtain reliable distribution of $\rm{DM_{host}}$ at the low DM end from observations, We also run the fitting in different range. The fitting results will change slightly if the fitting range is $\rm{DM_{host}}> 10 \, \rm{pc \, cm^{-3}}$, and change moderately if the range is $\rm{DM_{host}}> 30 \, \rm{pc \, cm^{-3}}$ (see Table \ref{tab:fit range test} for details). Visually, only the distribution of $\rm{DM_{host}}$ with `PSFR-VNum' model for galaxies in the TNG simulation can likely be  described by the log-normal formula, others deviate from the log-normal distribution, especially at the low DM end, $<10 \rm{pc \ cm^{-3}}$. We further test the normality of $\rm{ln(DM_{host})}$ using the Shapiro-Wilk test by python package `scipy.stats.shapiro'. Note, the results of such test of normality may not be accurate, if the size of sample is larger than 5000. 
 
Meanwhile, as section \ref{subsec:observation compare} will show, a few thousands of events could offer reliable statistics. Therefore, we randomly select 1000 events from the mock FRB samples of each model to carry the Shapiro-Wilk test. We repeat this procedure 20 times for each model and the mean p-value are listed in Table \ref{tab:fit param mainly}. The Shapiro-Wilk tests suggest that $\rm{DM_{host}}$ deviate from the log-normal distribution for all the two population models. In contrast, the distribution of $\rm{DM_{host, ISM}}$ in \cite{2020ApJ...900..170Z} for model of non-repeating events is reported to be well fitted by the log-normal function.

On the other hand, Figure \ref{fig:compare model in two simu} suggests that $\rm{DM_{host, halo}}$ is an non-negligible component of $\rm{DM_{host}}$ in the TNG samples, but plays minor role in the Illustris samples. Note that we take the gas and electron within the range of 5$\rm{R}_{50}$ as host galaxy/subhalo term, which is likely to be reasonable for the TNG samples, but is overestimated for galaxies in Illustris. The galaxy size in Illustris is generally larger than the observation (e.g. \citealt{2018MNRAS.473.4077P}). As will be demonstrated in the following sections, about 1\%-6\% FRB sources are locating outside the radius 5$\rm{R}_{50}$ for both simulations. These events would be considered to happen in the halos of host galaxy. For these events, the value of $\rm{DM_{host}}$ are equal to $\rm{DM_{host,halo}}$ in our definition. Excluding these extremely cases, we find that the median ratio of $\rm{DM_{host, halo} / DM_{host}}$ are 17.0\%, 40.1\% (3.7\%, 6.8\%) with the `PSFR-VNum' and `PStar-VNum' models respectively for the TNG (Illustris) galaxies. If we don't exclude events outside of 5$\rm{R}_{50}$, these fractions will increase slightly.

So far, the observational estimation on $\rm{DM_{halo, ISM}}$ and $\rm{DM_{host, halo}}$ are available only for a coupe of localized events. \cite{2021ApJ...922..173C} give the measurements of $\rm{ DM_{host, ISM}=82\pm35 \, pc \, cm^{-3}}$ and $\rm{ DM_{host, halo}=55\pm25 \, pc \, cm^{-3}}$ for FRB 190608. Meanwhile, $\rm{DM_{host,halo}}$ of FRB 20200120E overwhelmingly dominates over $\rm{DM_{Host,ISM}}$, which is negligible due to the event's position in the associated globular cluster and its offset from M81 center \citep[][]{2021arXiv210511445K}. Therefore, the CGM in host halo is an important part of $\rm{DM_{host}}$ for these two events, in consistent with our calculation for the TNG samples. However, some previous works didn't include the contribution from the CGM in host halo, and only account for $\rm{DM_{host, ISM}}$ when they estimate $\rm{DM_{host}}$ \citep[e.g.][]{2020ApJ...900..170Z}. Therefore, their values of $\rm{DM_{host}}$ are most likely underestimated. The redshift/distance of un-localized FRBs are usually inferred by the $\rm{DM_{IGM}} - z$ relation \citep[e.g.][]{2003ApJ...598L..79I,2004MNRAS.348..999I,2014ApJ...780L..33M}. An underestimated $\rm{DM_{host}}$ will lead to an overestimated redshift. This effect would be more serious for those FRB with relatively smaller values of $\rm{DM_{IGM}}$. 

\begin{figure}
\begin{center}
	\includegraphics[width=0.70\columnwidth]{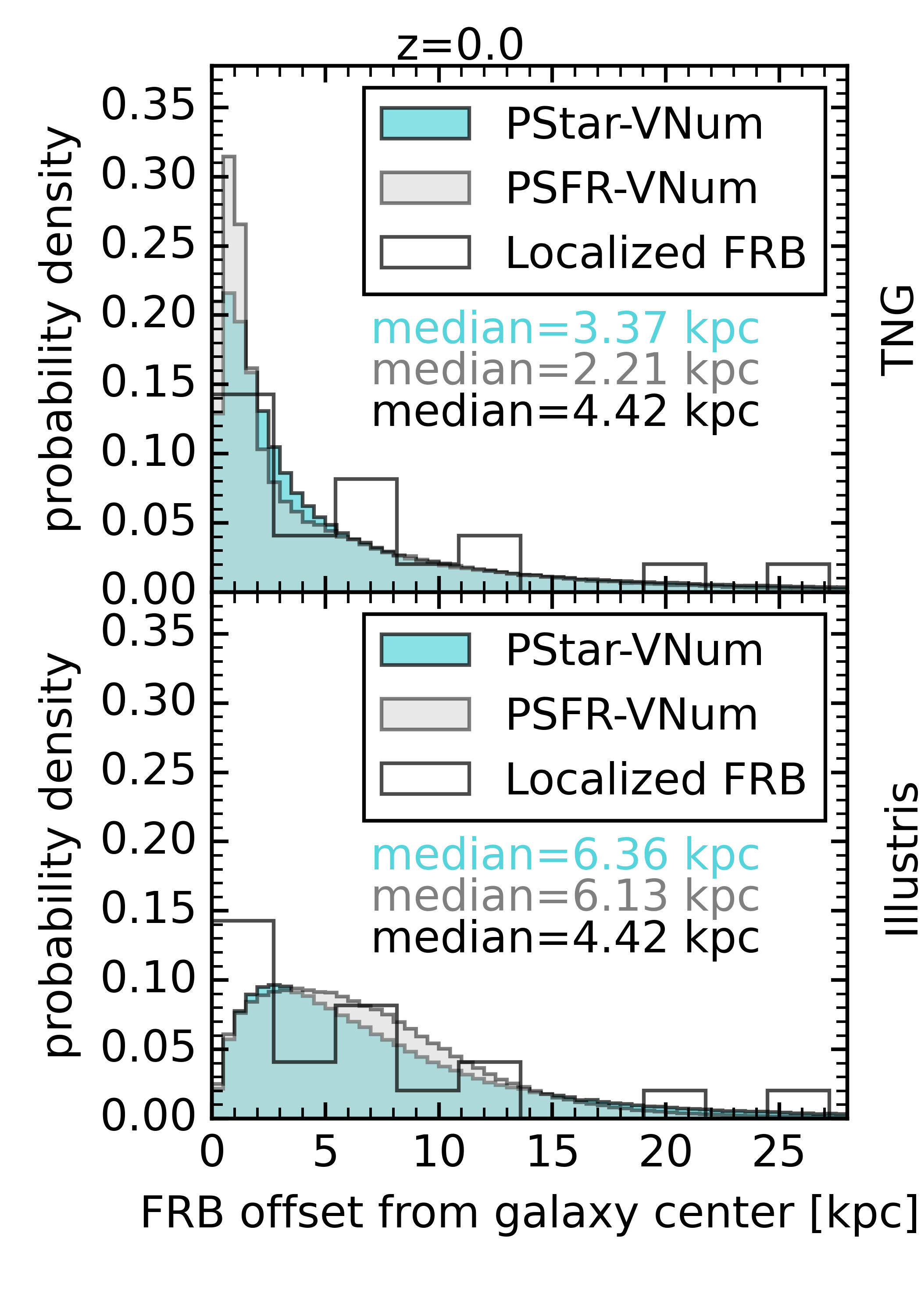}
    \caption{The distribution of the projected offset distance of FRB from galaxy center. First (second) panel: the results based on TNG100-1 (Illustris-1) simulation. The blue, gray histograms as well as texts show the distribution and median value of 'PStar-VNum' and 'PSFR-VNum' model, respectively. The black histogram and text are the distribution and median value of 19 localized FRB. }
    \label{fig:FRB distance distriution}
	\end{center}
\end{figure}

\begin{figure}
\begin{center}

	\includegraphics[width=0.70\columnwidth]{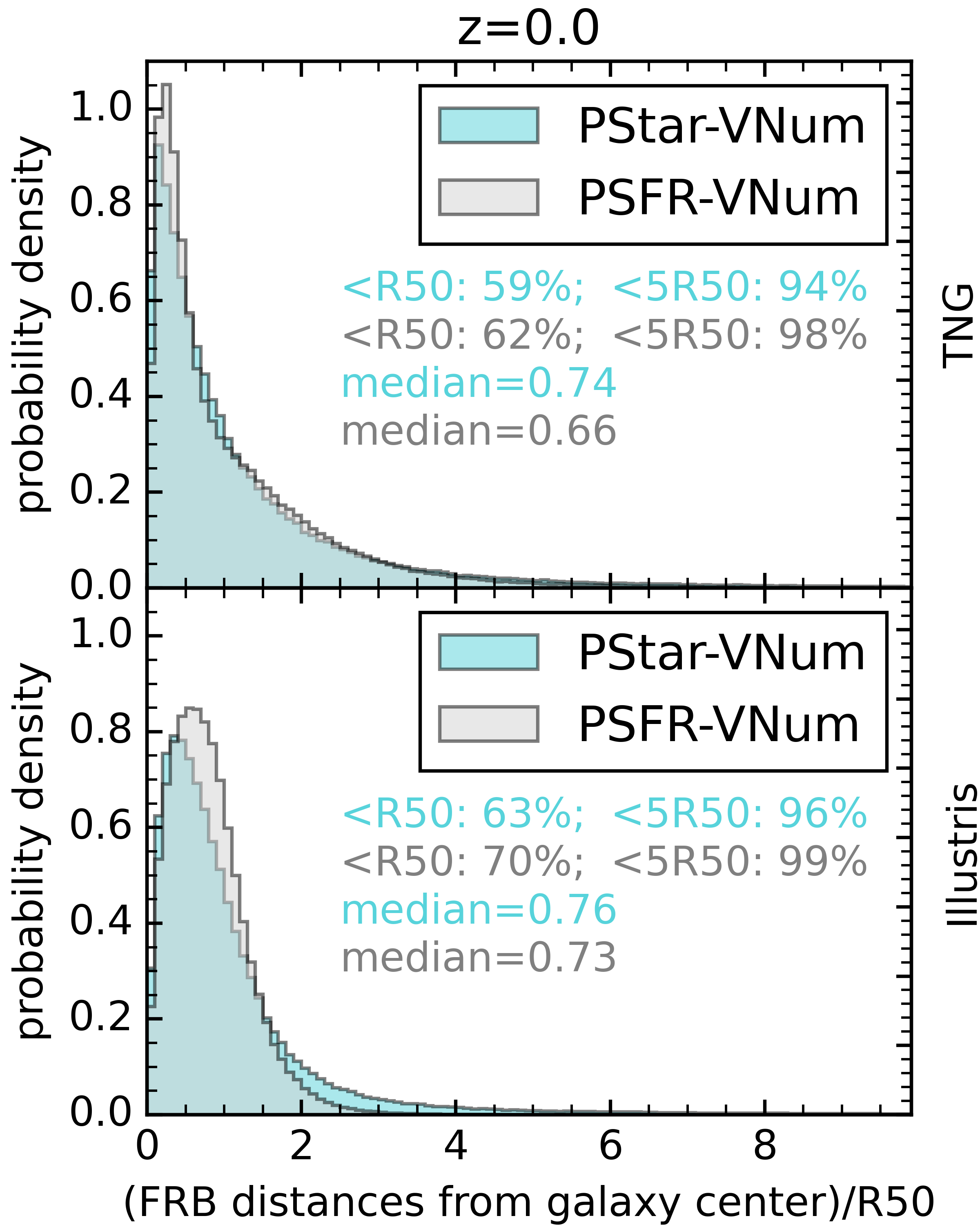}
    \caption{Similar to Figure \ref{fig:FRB distance distriution}, but the projected offset distance is scaled with the half stellar radius R50.}
    \label{fig:FRB normalized d distribution}
    \end{center}
\end{figure}

\subsection{Causes of differences between population models and simulations}

\begin{figure*}
	\includegraphics[width=2.0\columnwidth]{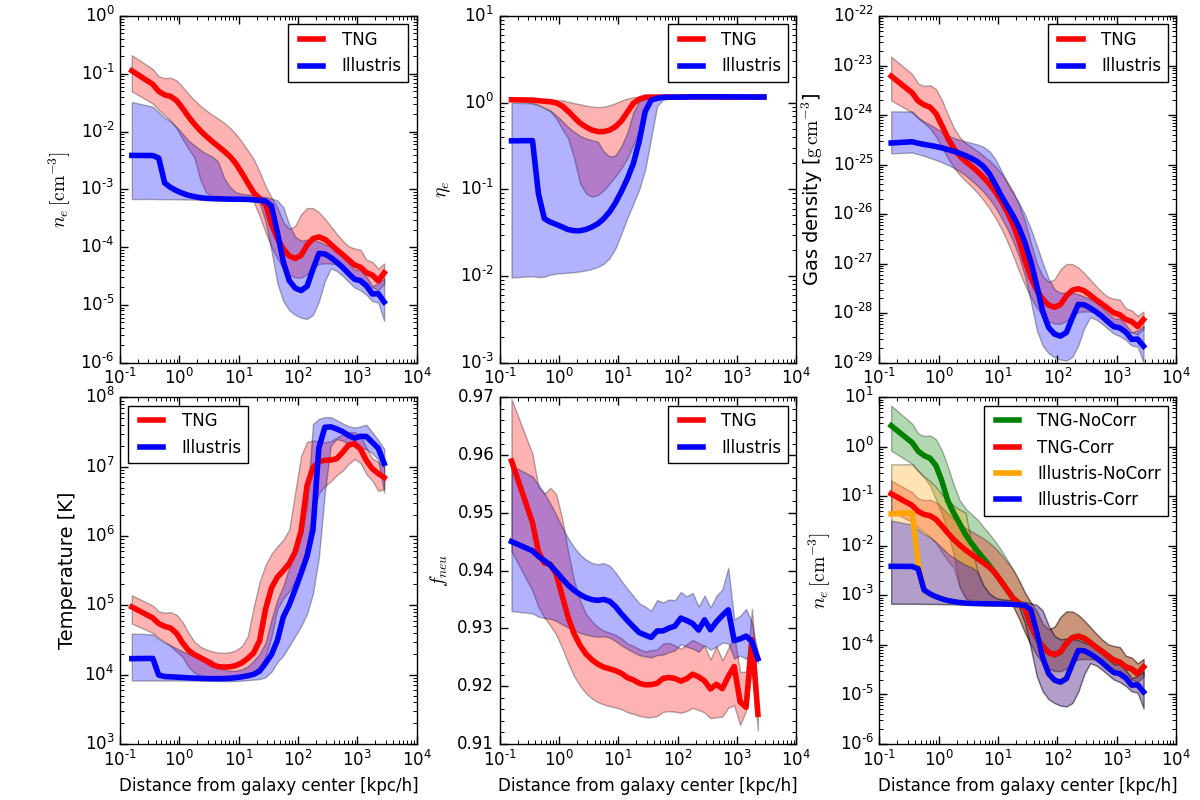}
    \caption{The electron density (top left), electron abundance (top middle), gas density (top right), temperature (bottom left), neutral fraction (bottom middle) as a function of distance from galaxy center in TNG100-1 (red) and Illustris-1 (blue) simulations. The bottom right panel shows the impact of dividing the star forming gas cells into neutral and ionized components (correction, Corr for short) or not (NoCorr). The solid line indicate the median value of all the galaxies. The upper and lower limit of shadow region show the 25 and 75 percentile. }
    \label{fig:ne_rho_eabu_profile}
\end{figure*}

The differences of DM that contributed by host galaxy and halo between the two population models must primarily caused by the differences in the spatial distribution of FRB events. To make a direct comparison with current and future observations, we probe the projected offset distances of FRB from the galaxy center. We project the 3D spatial distribution of FRB events and their host galaxy center onto the 'yz' plane of Cartesian coordinates in the simulations, and measure the projected offset distance $\rm{R_{off}}$. We also have tested the projection onto other two planes, which has minor impact on our results. Figure \ref{fig:FRB distance distriution} shows the projected offset distances of mock FRB events from the center of host galaxy. The distribution and median value of $\rm{R_{off}}$ indicates that the events produced by the 'PSFR-VNum' model tend to locate at places more close to the galaxy center than the events produced by the 'PStar-VNum' model. Therefore the line-of-sights to FRBs will penetrate more deep into the central region of host galaxies, if FRBs trace the star formation rate. This could explain that the values of $\rm{DM_{host, ISM}}$ and $\rm{DM_{host}}$ in the 'PSFR-VNum' model are relatively larger than those in the 'PStar-VNum' model, while $\rm{DM_{host, halo}}$ is close in two models. 
For galaxies in the TNG simulation, the mock FRB events mostly occur within 5kpc from the center of host, the number of events decreases with the offset distance. In contrast, the number of mock FRB events in the Illustris samples firstly increases with $\rm{R_{off}}$, peaks at $\rm{R_{off}}=2-5$ kpc, and then declines with increasing $\rm{R_{off}}$. The median value of $\rm{R_{off}}$ in the Illustris sample is around two times of that in the TNG sample, which can be attributed to the larger galaxy sizes in Illustris than in TNG \citep[][]{2018MNRAS.473.4077P}. In all the models, there is a tiny fraction of mock FRB events locating relatively far from the center of host, e.g. a projected distance larger than 25 kpc. This feature agrees with \cite{2020AcA....70...87J}, which also suggests a small fraction of events would have $\rm{R_{off}}>25 kpc$. 

\begin{figure}
    \hspace{-0.2cm}
	\includegraphics[width=1.05\columnwidth]{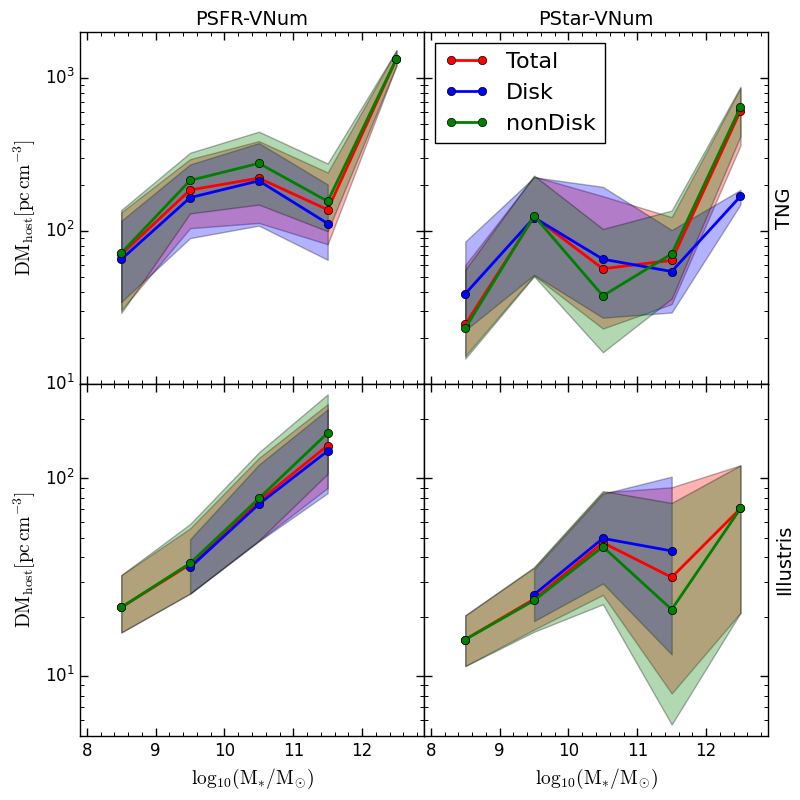}
    \caption{The left, right column indicate $\rm{DM_{Host}}$ of galaxies with 'PSFR-VNum'  and 'PStar-VNum' model respectively in different stellar mass bin at z=0. The top (bottom) row shows results based on galaxies samples in the TNG100-1 (Illustris-1) simulations. The red, blue, green lines show the results of all the galaxies samples, disk and non-disk samples respectively. The lower and upper limit of shadow region show the 25 and 75 percentile.}
    \label{fig:different model massbin z=0}
\end{figure}

Comparing the offset distances of mock events with observations could provide hints on the physical origin of FRBs (\citealt{2019ApJ...886..110M}). However, there are only a handful of localised events, which confines the power of such comparison at present. Here, we carry out a tentative trial. The redshifts of the 19 localized FRB are mainly lower than 0.5. We find that for galaxies in the two FRB population models, there is little difference of the distribution of projected offset distances from galaxy center between $z=0$ and $z=0.5$, as shown by Figure \ref{fig:distance distribution z0.5}. Thus, we present the offsets of the 19 localized FRBs in Figure \ref{fig:FRB distance distriution}, and make a comparison with mock events. While suffering from significant fluctuations, the number of localised events display a trend to decline with $\rm{R_{off}}$, which can be better explained by the galaxy sample extracted from the TNG simulation. Moreover, the median $\rm{R_{off}}$ of localised events is more close to the result with the TNG samples. On the other hand, it seems the 'PStar-VNum' model predicts a median $\rm{R_{off}}$ more close to the observed events.  As more and more FRBs will be localized in the future, it would enable us to make more precise restrictions on FRB models and feedback models in simulations according to the spatial offset from galaxy center. 

We further probe the spatial distribution of mock FRBs in galaxies by the offset distances scaled by the half stellar radius, which is illustrated by Figure \ref{fig:FRB normalized d distribution}. FRB events are mainly distributed within $R_{50}$ from the center of galaxy. For galaxies in the TNG simulation, 98\%, and 94\% of the FRB sources placed by the PSFR-VNum, and PStar-VNum model are within the radius of 5$R_{50}$. The corresponding fractions of mock FRB placed in Illustris galaxies are moderately higher. 

We now discuss the difference between results based on galaxies samples from the two simulations - TNG100-1 and Illustris-1. For both of the two distinct FRB population models, the values of DM components measured in the Illustris galaxy samples are generally much smaller than that in the TNG samples, resulting in much lower median values. We found that this discrepancy results from the combining effects of electron distribution and the central offset of FRB, $\rm{R_{off}}$. The top left panel in Figure \ref{fig:ne_rho_eabu_profile} shows the electron number density profile $n_e$ in the host galaxy samples selected from the two simulations at $z=0$. The electron number density in the central region ($r<=1$ kpc) of the TNG galaxy samples is higher than that in the Illustris galaxies by over an order of magnitude. The difference narrows gradually outward, and becomes marginally visible at $r\sim 10-15$ kpc. At $r\geq 10-15$ kpc, $n_e$ drops sharply in the galaxies selected from both simulations. The discrepancy of $n_e$ within $r\leq 10$ kpc is much stronger than gas density, and should be mainly attributed to the very different levels of electron abundance $\eta_e$, as shown by the top middle and right panels in Figure \ref{fig:ne_rho_eabu_profile}. We find that the much higher $\eta_e$ in the central region of the TNG galaxy samples results from a hotter gas temperature in comparison with the Illustris samples. The reason for the difference of the gas temperature in the central regions of galaxies between the two simulations should be related to the distinct feedback models. \citep[][]{2017MNRAS.465.3291W, 2018MNRAS.473.4077P}. 

It is noted that, as described in Section \ref{sub:dm_calc}, there are both non-star-forming and star forming gas cells in the galaxies. When counting the electrons in cells, the star forming cells are divided into two portions, i.e. the neutral and ionized, as a `correction'. The star-forming cells occupy $\sim 21\% $, and $8\%$ of all the gas cells in the galaxy samples selected from the TNG and Illustris simulations. Meanwhile, the fraction of neutral gas in the star-forming cells is close in these two simulations, $\sim 91\%-93\%$. The effect of the `correction' for the star forming gas cells on the electron density is demonstrated by the bottom right panel of Figure \ref{fig:ne_rho_eabu_profile}. For galaxies in the TNG simulation, this correction could avoid an overestimated electron density in the central region within $r\leq 1$ kpc. Yet, the correction has moderate effect on the results obtained from the selected galaxies in the Illustris simulation.

\subsection{dependence on galaxy stellar mass}
\label{sub:morphology and stellar mass effect}

The DM caused by host galaxies and halos is expected to depend on the stellar mass of host galaxy. Figure \ref{fig:different model massbin z=0} shows the median value, and the 25 and 75 percentiles of $\rm{DM_{host}}$ as a function of the stellar mass of hosts at $z=0$. The probability distribution of $\rm{DM_{host, ISM}}$, $\rm{DM_{host, halo}}$, $\rm{DM_{host}}$ of the total galaxies and their median values in different stellar mass bins at $z=0$ are displayed in Figure \ref{fig:tng massbin pdf of two models with 5R50}. The differences of $\rm{DM_{host}}$ between the two population models, and between the two simulations are similar in different mass bins. Figure \ref{fig:different model massbin z=0} indicates that $\rm{DM_{host}}$ generally increases with stellar mass at $M_*\leq 10^{10.5} M_{\odot}$ and shows notable fluctuations at $M_* >10^{10.5} M_{\odot}$ for the TNG samples, in good agreement with the trend in \cite{2020AcA....70...87J}. For galaxies in the Illustris simulation, $\rm{DM_{host}}$ increases with stellar mass in the whole mass range of the PSFR model, but also shows fluctuations at $M_* >10^{10.5} M_{\odot}$ with the PStar model. The probable reasons are as follows. For a host galaxy with a stellar mass of $M_*\leq 10^{10.5} M_{\odot}$, its gas mass usually increases with $M_*$. Meanwhile, the mass and temperature of CGM in the parent dark matter halo also increases with $M_*$. However, for galaxies with stellar mass greater than $M_*=10^{10.5} M_{\odot}$, their gas fractions in the central region may decrease due to AGN feedback (\citealt{2018MNRAS.475..624N}), yet the number of electrons in their parent halo still increases with $M_*$. The fraction of $\rm{DM_{host}}$ contributed from the CGM in halo, i.e., $\rm{DM_{host, halo}}$, generally increases with the stellar mass of host galaxy, which can be inferred from Figure \ref{fig:tng massbin pdf of two models with 5R50} . In addition, the fluctuations of $\rm{DM_{host}}$ at the high stellar mass end should partly result from the limited number of samples. 

\begin{figure}
    \hspace{-0.2cm}
    \includegraphics[width=1.00\columnwidth]{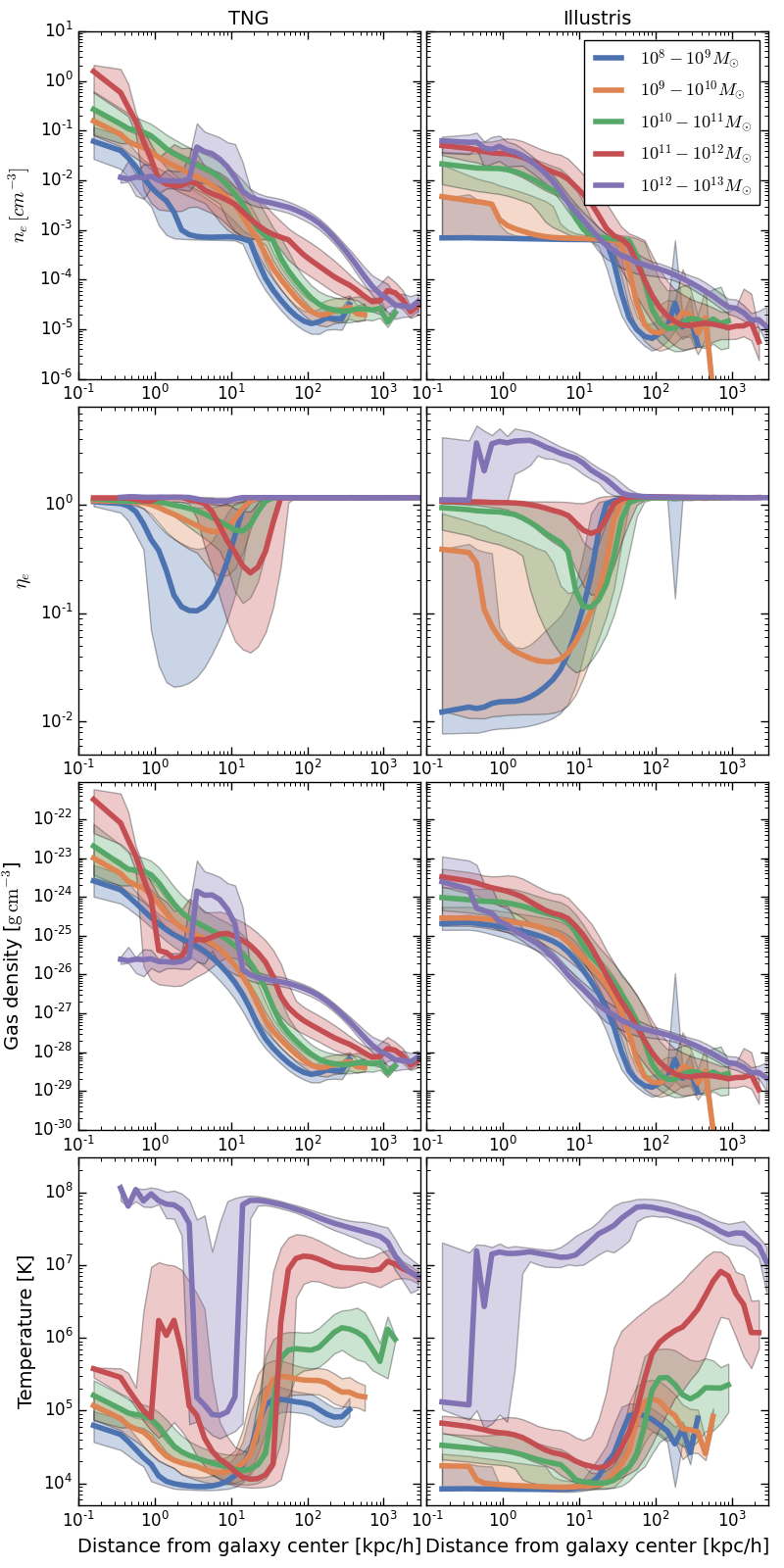}
    \caption{The first, second, third and fourth rows show the electron density, electron abundance, gas density, temperature as a function of distance from galaxy center respectively in the TNG100-1 (left) and Illustris-1 (right) simulations. The  blue, orange, green, red, purple solid line indicate the median value of all the galaxies with $10^8-10^9 M_\odot, \ 10^{9}-10^{10} M_\odot, \ 10^{10}-10^{11} M_\odot, \ 10^{11}-10^{12} M_\odot, \ 10^{12}-10^{13} M_\odot $ stellar mass respectively. The upper and lower limit of shadow region indicate the 25 and 75 percentile. }
    \label{fig:compare massbin profile in tng}
\end{figure}

We can see that the fluctuations of $\rm{DM_{host}}$ at the high stellar mass end are more significant for galaxies in the TNG simulation. Meanwhile, the `PStar' model exhibits stronger fluctuation, which is very likely related to a more extended spatial distribution of mock events. To explore what causes the stronger fluctuations in the TNG sample, we trace the $n_e$ profiles in both the TNG and Illustris samples, as shown in Figure \ref{fig:compare massbin profile in tng} respectively. We find that the electron densities in the central region $r \lesssim 3 \rm{kpc}/h $ of galaxies with $M_* \geq 10^{12-13}\,M_{\odot}$ are lower than those with $M_* \leq 10^{9-12}\,M_{\odot}$ in the TNG samples. Meanwhile, the electron density in galaxies with $M_*=10^{11-12}\,M_{\odot}$ drops sharply at $r\simeq 0.5 \rm{kpc}/h $ and is lower than that of galaxies with $M_*=10^{9-11}\,M_{\odot}$ in the range $0.8\,\rm{kpc}/h \lesssim r \lesssim 10 \rm{kpc}/h $.  In contrast, this turnover is not found in the Illustris samples. The probable reason is that the active galactic nucleus (AGN) feedback is more effective in the TNG simulation where both kinetic and thermal feedback mechanisms are included. In comparison, only the thermal channel is implemented in Illustris. More effective AGN feedback in the TNG simulation could expel the gas in central region outward more easily, as shown by the gas density plots in Figure \ref{fig:compare massbin profile in tng}. Note that, AGN feedback usually works in galaxies more massive than $M_* \sim 10^{10}\,M_{\odot}$. Therefore, galaxies in the TNG simulation with $M_* =10^{11-12}\,M_{\odot}$ may have been affected by this effect more fiercely. 

Again, different feedback models in simulations and different FRB population models would lead to considerable differences in $\rm{DM_{host}}$. Therefore, if we have a better understanding of the nature and population of FRB sources, and a larger number of localized FRBs and the information of their $\rm{DM_{host}}$, we could in turn constrain the models of feedback which shape the physical properties of galaxies.

\begin{table}
	\centering
	\caption{The median value of $\rm{DM_{host}}$ of distribution that based on galaxies samples in the TNG100-1 and Illustris-1 simulations by different FRB population models. }
	\label{tab:fit param disk nondisk}
	\begin{tabular}{m{4em} m{7em} m{6em} m{4em} } 
		\hline
		simulation & model & type of galaxies & median [$\rm{pc\,cm^{-3}}$]   \\
		\hline
		\multirow{4}{4em}{TNG} & \multirow{2}{7em}{PSFR-VNum}  & Disk  & 182.97   \\
		                                  &           & NonDisk  & 172.07 \\
		                        
		                        & \multirow{2}{7em}{PStar-VNum} & Disk  & 69.41 \\
		                                              &    & NonDisk & 55.90 \\
		\hline
		\multirow{4}{4em}{Illustris} & \multirow{2}{7em}{PSFR-VNum} & Disk & 62.67 \\
		                                            &           & NonDisk & 47.57 \\

		                        & \multirow{2}{7em}{PStar-VNum} & Disk  & 42.79 \\
		                                              &    & NonDisk & 26.27 \\

		\hline
	\end{tabular}
\end{table}

\begin{figure*}
	\includegraphics[width=1.8\columnwidth]{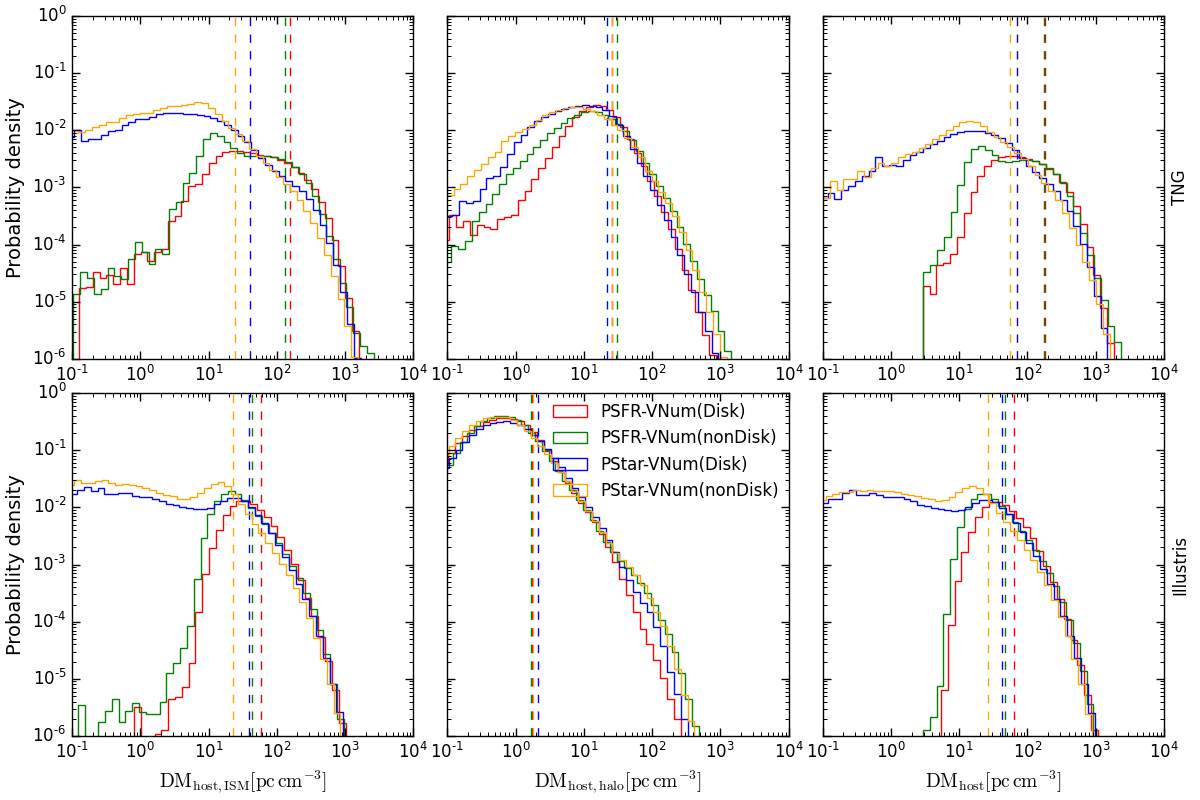}
    \caption{
    The probability distribution of $\rm{DM_{host}}$ for disc and non-disc galaxies at $z=0.0$ . The left, middle and right column are distributions of DM caused by the 'ISM', 'Halo' and 'Host' components respectively. Red, blue, green and orange histogram (vertical dashed lines) are distribution (median) of DM for the 'PSFR-VNum' and 'PStar-VNum'  models with galaxies in the TNG100-1 and Illustris-1 respectively.}
    \label{fig:compare DiskNondisk}
\end{figure*}
\subsection{disc and non-disc hosts}
Most of the 19 localized FRBs are hosted by disk galaxies, while a few are hosted by elliptical galaxies (\citealt{2020ApJ...903..152H}). It would be worth investigating whether $\rm{DM_{host}}$ is related to galaxy morphology or not. We divide the galaxy samples into two subgroups, disk and non-disk galaxies, and identify disk galaxies from our samples following the method in \cite{2020ApJ...895...92Z}. Table \ref{tab:num galaxies and FRBs and in disk} presents the number of identified disk galaxies. There is significant difference in the fraction of disk samples between the TNG and Illustris simulations. We find that the overall distribution of $\rm{DM_{host, ISM}}$, $\rm{DM_{host, halo}}$ and $\rm{DM_{host}}$ are similar between disk and non-disk galaxies in both simulations. Yet, we should remind the readers that non-disk galaxies include elliptical and irregulars. There might be some differences in $\rm{DM_{host}}$ between elliptical and irregulars. However, splitting these two types of galaxies is not a easy task, and we skip it in this work.

The distribution of $\rm{DM_{host}}$ of disk and non-disk galaxies in the two simulations are shown in Figure \ref{fig:compare DiskNondisk}. The median values of disk galaxies sample are higher by $\sim 10-30\%$ for both population model, and for both simulations, which could be due to a combining effect of the mass function and the gas content of host galaxy. Table \ref{tab:fit param disk nondisk} lists the median value.

\begin{figure}
	\includegraphics[width=1.05\columnwidth]{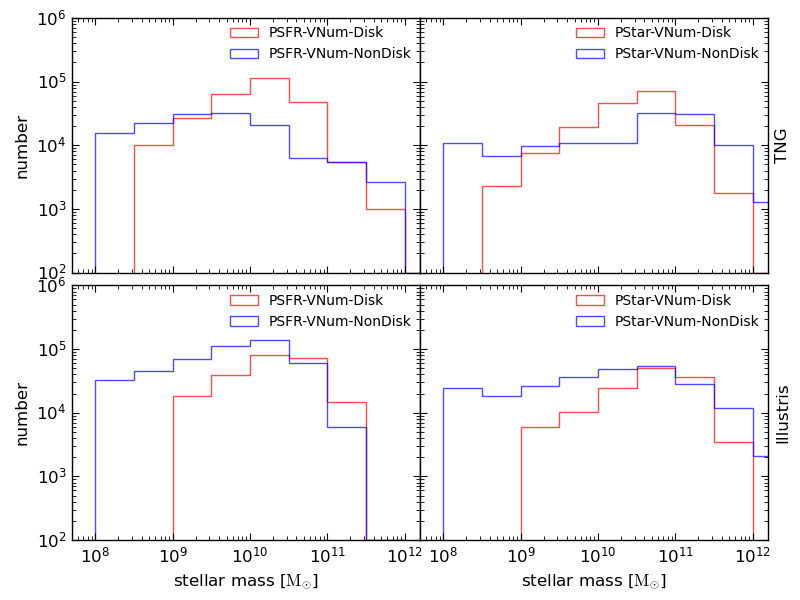}
    \caption{The stellar mass function of host galaxies weighted by the number of mock FRB events. The red (blue) shows results of PSFR-VNum/PStar-VNum population model with disk (non-disk) galaxies. Top (Bottom): the results based on galaxies in TNG100-1 (Illustris-1).}
    \label{fig:MassFunction of Disk and nonDisk}
\end{figure}

Figure \ref{fig:different model massbin z=0} also shows that for PSFR-VNum model the median value of $\rm{DM_{host}}$ of non-disk samples in each stellar mass bin is slightly higher than disk samples in both the TNG and Illustris simulations. The PStar-VNum model exhibits an opposite trend when stellar mass $M_* \leq 10^{10.5} \rm{M_{\odot}}$. Nevertheless, the median value of $\rm{DM_{host}}$ of the whole disk samples is higher than that of non-disk as mentioned above, which seems to contradict with the relative strength in each mass bin for the PSFR-VNum model. This feature is mainly caused by the difference in the distribution of FRB number as a function of host stellar mass between disk and non-disc samples, which is shown by Figure \ref{fig:MassFunction of Disk and nonDisk}. We can see that the median stellar mass of hosts is moderately larger than $10^{10} \rm{M_{\odot}}$ for mock events that placed in disk galaxies, which is larger than that in non-disc galaxies for both population model and both simulations. As $\rm{DM_{host}}$ generally increases with the hosts' stellar mass when $M_* \leq 10^{10.5} \rm{M_{\odot}}$, it is reasonable to find a larger median of $\rm{DM_{host}}$ for the disk samples than the non-disc samples.

To provide a more direct comparison with previous studies, we look for disk galaxies with stellar mass $10^{10} - 10^{11} M_{\odot}$, i.e., the Milky-Way-like galaxies. The median value of $\rm{DM_{host, ISM}, DM_{host, halo}, DM_{host}}$ for MW-like hosts are listed in Table \ref{tab:DM of MW-like galaxies}. The corresponding values are $\rm{180.66,\, 30.10,\, 212.93\, pc \, cm^{-3}}$ for the PSFR-VNum population models with the TNG galaxy samples. And the mean $\rm{DM_{host, halo}}$ for MW-like hosts is $\rm{44 \, pc\,cm^{-3}}$. In comparison, the estimated $\rm{DM_{MW, halo}}$ is 50-80 $\rm{pc \, cm^{-3}}$ in \cite{2019MNRAS.485..648P}, and ranges from 30-245 $\rm{pc \, cm^{-3}}$ with a mean value of 43 $\rm{pc \, cm^{-3}}$ in \cite{2020ApJ...888..105Y}. The discrepancy would be partly attributed to the extended size of ISM region, i.e., 5$R_{50}$, which are adopted to have larger values, e.g. $\simeq 24-46$ kpc depending on the models, than the corresponding size ($\simeq 20\, $kpc) in those literature. 
 
\begin{table}
	\centering
	\caption{The median value of $\rm{DM_{host} \, [pc\,cm^{-3}]} $ for MW-like galaxies in the TNG100-1 and Illustris-1 simulations at $z=0$. }
	\label{tab:DM of MW-like galaxies}
	\begin{tabular}{m{4em} m{3em} m{4em} m{4em} m{4em}  } 
		\hline
		simulation & model & $\rm{DM_{host,ISM}}$  & $\rm{DM_{host,halo}}$  & $\rm{DM_{host}}$  \\
		\hline
		\multirow{3}{3em}{TNG}  & PSFR-VNum & 180.66 & 30.10 & 212.93    \\

		                        & PStar-VNum   & 36.26 & 22.92 & 65.45 \\

		\hline
		\multirow{3}{3em}{Illustris} & PSFR-VNum   & 70.33 & 1.96 & 74.17  \\
		                        
		                        & PStar-VNum  & 46.08 & 2.31 & 49.80   \\

		\hline
	\end{tabular}
\end{table}

\subsection{redshift evolution}
\begin{figure}
	\includegraphics[width=\columnwidth]{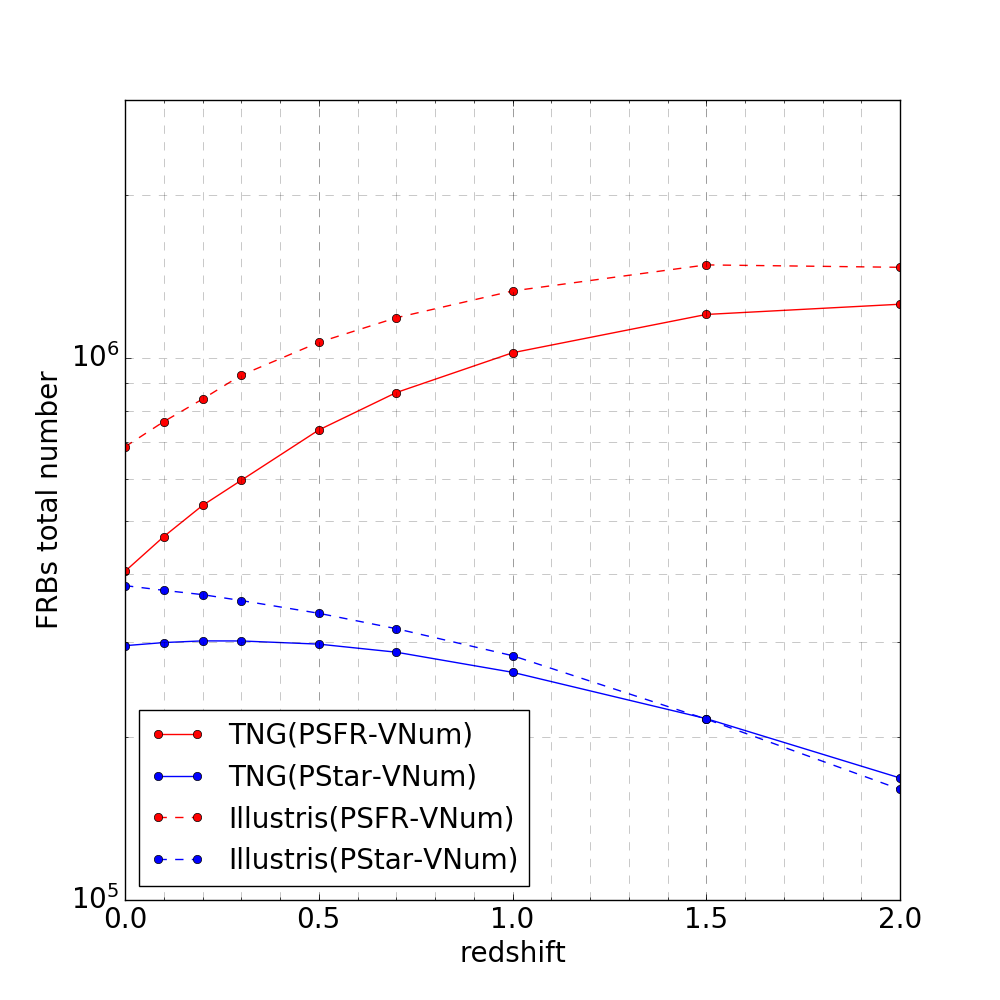}
    \caption{The red and blue lines show the evolution of total Number of mock FRB event with 'PSFR-VNum' and 'PStar-VNum' models respectively between redshift $0.0-2.0$. The solid and dashed lines indicate results based on TNG100-1 and Illustris-1 simulation respectively.}
    \label{fig:redshift evolution of NumFRB}
\end{figure}

We extend our analysis to higher redshifts in this subsection. Figure \ref{fig:redshift evolution of NumFRB} compares the number of mock FRB events according to eqn. \ref{eqn:number FRB of Mstar} and \ref{eqn:number FRB of SFR} in the different models between redshift $z=0$ and $z=2$. The number of mock events increases with redshifts in the `PSFR' population model, and conversely, decreases with redshifts in the `PStar' model, as expected. In the former, the total number of mock FRBs is above 1 million at $z=2$, which is higher than the latter model by an order of magnitude. In addition, we also randomly draw 20 line-of-sights for each mock FRB events at $z>0$, which is in consistent with the treatment at $z=0$.

\begin{figure}
	\includegraphics[width=\columnwidth]{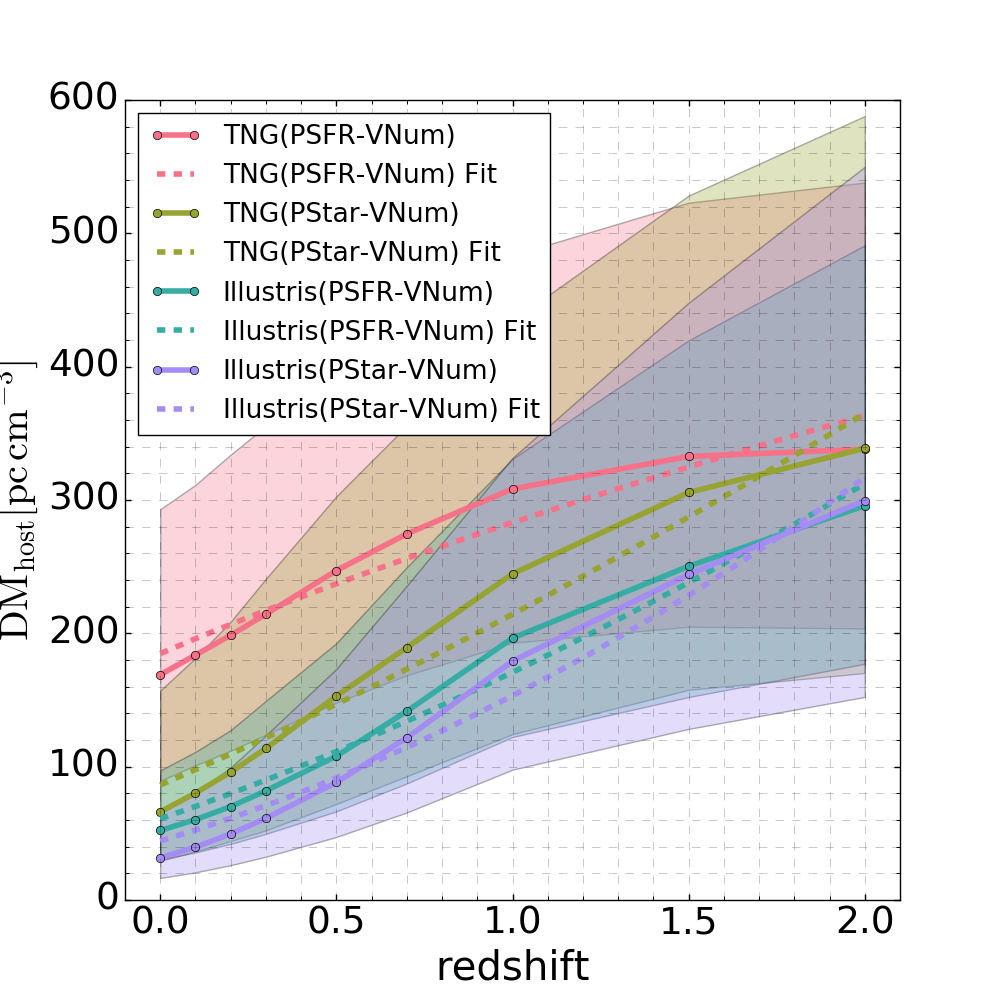}
    \caption{The red and brown (blue and purple) solid lines show the median value of $\rm{DM_{Host}} $ as a function of redshift with 'PSFR-VNum' and 'PStar-VNum' model respectively for galaxies in TNG100-1 (Illustris-1). The dashed lines are the fitting results according to equation \ref{eqn:DM_Host with z}. The lower and upper limit of shadow region show the 25 and 75 percentile.}
    \label{fig:redshift evolution of model}
\end{figure}

\begin{figure}
	\includegraphics[width=\columnwidth]{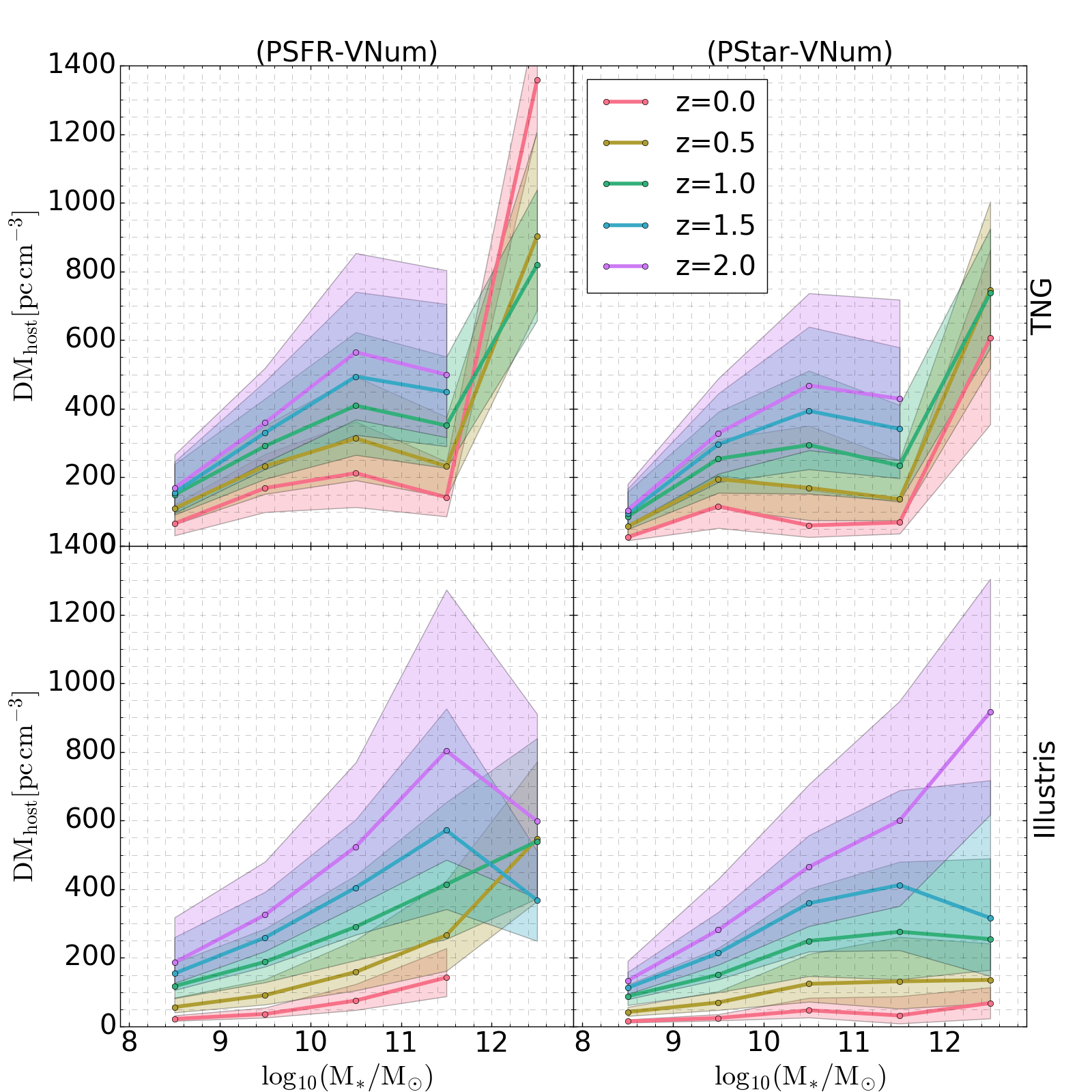}
    \caption{The red, brown, green, blue and purple lines show $\rm{DM_{host}}$ of galaxies in different stellar mass bins at redshift z=0.0, 0.5, 1.0, 1.5, 2.0 respectively. The lower and upper limit of shadow region show the 25 and 75 percentile.}
    \label{fig:redshift evolution in massbin}
\end{figure}

Figure \ref{fig:redshift evolution of model} displays the median, 75 and 25 percentiles of the total DM contributed by the host galaxy and halo, $\rm{DM_{host}}$, as a function of redshifts in both the population models, i.e. PSFR-VNum and PStar-VNum, based on galaxies samples in the TNG100-1 and Illustris-1 simulations. $\rm{DM_{host}}$ increases with increasing redshifts, which is consistent with \cite{2020AcA....70...87J} and \cite{2020ApJ...900..170Z}. The primary reason is that the physical densities of gas and electron in galaxy and halo increase with redshifts. An empirical equation 
\begin{equation}
    \rm{DM_{host}(z) = A(1+z)^{\alpha}}
\label{eqn:DM_Host with z}
\end{equation}
is implemented to fit the evolution of $\rm{DM_{host}}$ with redshifts, following \cite{2020ApJ...900..170Z}. The fitting parameters are listed in Table \ref{tab:fit DM redshift evolution}. The measured index $\alpha$ in our sample differs from \cite{2020ApJ...900..170Z} moderately. 

The median value of $\rm{DM_{host}}$ in the TNG simulation is larger than that in the Illustris simulation at all redshifts. $\rm{DM_{host}}$ of galaxies in the TNG simulation with the PStar-VNum model can even be larger than that of galaxies in the Illustris with PSFR-VNum model. For galaxies selected from the two simulations, the PSFR-VNum model predicts a larger $\rm{DM_{host}}$ than that of the PStar-VNum model's at high z. The discrepancy between two population models narrows with increasing redshifts, and becomes negligible at $z=2$. As the redshifts increase, the distributions of FRB offset from galaxy center in these two population models become more similar, as shown in Figure \ref{fig:distance distribution z0.5}. It indicates that the distribution of stellar mass and star-forming cells are more consistent with each other at higher redshifts. This is reasonable, as the cosmic star formation rates increase gradually from $z=0$ toward higher redshifts and peaks at $z \sim 2$. Hence, the stellar mass at higher redshifts will be more closely related to newly born stars.

Figure \ref{fig:redshift evolution in massbin} shows the redshift evolution of the median, 75 and 25 percentiles of $\rm{DM_{host}}$ in different stellar mass bins. The overall trend is similar in different mass bins, i.e., $\rm{DM_{host}}$ increasing with redshifts. Meanwhile, the dependence of $\rm{DM_{host}}$ on galaxy stellar mass at redshift $z>0$ is close to that at $z=0$ as discussed in section \ref{sub:morphology and stellar mass effect}. $\rm{DM_{host}}$ firstly increases with the stellar masses of host galaxies, and then exhibits fluctuations at the high mass end.


\begin{table}
	\centering
	\caption{The fitting parameters of $\rm{DM_{host}(z)}$ .}
	\label{tab:fit DM redshift evolution}
	\begin{tabular}{m{4em} m{6em} m{2em} m{2em} m{2em} m{2em} } 
		\hline
		simulation & model & A & $\sigma_{A}$ & $\alpha$ & $\sigma_{\alpha}$   \\
		\hline
		\multirow{3}{3em}{TNG}  & PSFR-VNum & 185 & 9 & 0.6 & 0.06    \\

		                        & PStar-VNum   & 86 & 8 & 1.3 & 0.1  \\

		\hline
		\multirow{3}{3em}{Illustris} & PSFR-VNum   & 61 & 5 & 1.5 & 0.1  \\
		                        
		                        & PStar-VNum  & 44 & 5 & 1.8 & 0.1   \\

		\hline
	\end{tabular}
\end{table}

\section{Discussions}
\label{sec:discussion}

\subsection{spatial coincident events}
\label{subsec:spatial coincident}

In our PSFR-VNum and PStar-VNum models, the probability that FRB events happen in a given cell (with a size of $\sim \rm{1-10 kpc}$) is proportional to its SFR and stellar mass respectively. In some rare cases, multiple FRB events can occur in the same cell that has a high SFR or large stellar mass. Therefore, if the spatial resolution of observation is insufficient, these sources might be falsely identified as repeating FRBs. We count the fraction of these spatial coincident events over all the mock FRB sources with the same projection method as mentioned in section 3.2, and the results are listed in Table \ref{tab:num samples of repeating}. For galaxies in the TNG100-1 and Illustris-1 simulations, the fractions of spatial coincident FRB in the PSFR-VNum model is $3.51\%$ and $1.89\%$ respectively, which are much higher than the corresponding fractions of $0.088\%$ and $0.063\%$ in the PStar-VNum model. The reason is that the distribution of FRB sources in the PSFR-VNum model tend to be more concentrated, and it would be easier to find multiple events in the same cell. Note, these fractions are measured by taking the spatial coincident events as non-repeating individual events. If spatial coincident events are falsely treated as the repeating FRBs, i.e., multiple non-repeating sources in the same cell are treated as one repeating source, the fractions will be $1.77\%$ and $\ 0.95\%$ for the PSFR-VNum model, and $\ 0.044\%$, $\ 0.032\% $ for the PStar-VNum model respectively. 

However, these fractions depend on the assumed number in a particular galaxy and the size of gas cells. For instance, if we lower down the event number in the PSFR-VNum model by changing the denominator in eqn. \ref{eqn:number FRB of SFR} from 0.033 to 0.33 $\rm{M_{\odot}/yr}$, the fraction of spatial coincident events will be reduced from $3.51\%$ to 0.25\% in TNG. Or if we change the denominator in eqn. \ref{eqn:number FRB of Mstar} from $10^9$ to $10^8 M_\odot$, the spatial coincident fraction will be increased from $0.088\%$ to $0.75\%$ for the PStar-VNum model of the TNG galaxies. We choose the denominator in eqn. \ref{eqn:number FRB of SFR} and \ref{eqn:number FRB of Mstar} to produce enough mock events to obtain reliable statistics on $\rm{DM_{host}}$ and to make the total number of events be comparable to \cite{2020ApJ...900..170Z}. In a more realistic treatment, any assumption of the number of mock events in a particular galaxy should take the time period into account. Namely, the event rate is a better indicator. 

Currently, a solid constraint on the event rate of FRB is not available, giving the uncertainty of physical origin. To provide a conceptual view on the event rate of FRBs in a particular galaxy, we look into the long-lived magnetars formed via merger of binary neutron stars. \cite{2019ApJ...886..110M} suggests that $\sim 3\%$ binary neutron stars (BNS) mergers could relate to FRB events. \cite{2019MNRAS.487.1675A} provides estimated relations between the number of BNS mergers per galaxy per unit time ($n_{\rm{BNS}}/\rm{Gyr}$), and the galaxy's stellar mass, star formation rate respectively in the EAGLE simulation. \cite{2022MNRAS.509.1557C} also estimates the relation of $n_{\rm{BNS}}/\rm{Gyr}-M_{*}$ based on samples in the TNG simulation, whose results are around the same order of magnitude with \cite{2019MNRAS.487.1675A}. For instance, a galaxy with stellar mass of $\rm{10^{10} M_{\odot}}$ (SFR=0.33 $\rm{M_{\odot}/yr}$) will expect about 19 (4) BNS mergers per million year. In comparison, we placed 11 mock FRBs in a galaxy with similar stellar mass, and SFR for the PStar-VNum and PSFR-VNum model respectively. A time period of tens of million years is needed in our models to fit with the rate of BNS mergers and the conversion ratio of BNS merges and FRBs in the literature. Consequently, the fractions of spatial coincident events discussed in the last two paragraphs would be much smaller if the observation period is tens of years. Yet, if the spatial resolution of observation is poorer than 1 kpc, the chance to observe spatial coincident events would increase.

\begin{table}
	\centering
	\caption{The number of spatial coincident FRB events in selected galaxies samples from the TNG100-1 and Illustris-1, and its fraction at $z=0$.}
	\label{tab:num samples of repeating}
	\begin{tabular}{m{4em} m{6em} m{8em} m{4em} } 
		\hline
		simulation & model & spatial coincident FRB events  &  fraction   \\
		\hline
		\multirow{3}{3em}{TNG}  & PSFR-VNum & 14213 & 3.51\%  \\

		                        & PStar-VNum & 260 & 0.088\%  \\

		\hline
		\multirow{3}{3em}{Illustris} & PSFR-VNum   & 12957 & 1.89\% \\
		                        
		                        & PStar-VNum   & 240 & 0.063\%   \\

		\hline
	\end{tabular}
\end{table}

Moreover, we set a upper limit of FRB number 100 in one galaxy as mentioned in the section \ref{sub:FRB population}. In fact, this is just for saving computational time, not a physical limitation. Taking into account this factor, we perform another run of $\rm{DM_{host}}$ calculation without placing the FRB number limit in each host galaxy, and find that this factor has little effect on the distribution of $\rm{DM_{host}}$ and its median value due to the small number of galaxies with either high stellar masses or higher SFRs. Meanwhile, it slightly increases the fraction of 'fake' repeating FRB.

Up to now, CHIME have detected 474 one-off and 62 repeat bursts from 18 repeaters \citep[][]{2021ApJS..257...59A} \footnote{\href{https://www.chime-frb.ca/catalog}{https://www.chime-frb.ca/catalog}}, so its fraction of repeaters is about 3.65\% which is close to the fraction of spatial coincident event in the PSFR-VNum population model based on galaxies samples in the TNG simulation. Although the fractions are close, we should be caution dealing with the repeaters in observation and spatial coincident events, which are likely to be different. The latter are caused by the spatial overlap effect, and their time lag would be much larger than observed repeaters. Using Monte Carlo simulations \cite{2021ApJ...906L...5A} found the true fractions of repeating FRB sources from the CHIME data, giving a prediction that the repeating FRB fractions should be less than 4\%. 

In the future, with more and more observational data, we may find some FRB events coming from the same spatial position with a relatively long time lag, but have significantly different properties such as $\rm{DM_{tot}}$ and scattering. Such FRB may be actually multiple FRB sources in the same spatial position, but the distribution of plasma in their local environment and nearby ISM can be different and hence lead to distinct feature on dispersion and scattering.

\subsection{Comparison to observations and applications}
\label{subsec:observation compare}

So far, some relatively rough estimations on the $\rm{DM_{host}}$ of several localised events have been reported, which ranges from 10 to 1121 $\rm{pc \ cm^{-3}}$. Currently, two approaches have been used to estimate $\rm{DM_{host}}$ of localised events. One is to subtract the foreground contribution from the total observed $\rm{DM_{obs}}$, e.g. FRB 121102 \citep[][]{2017ApJ...834L...7T}, FRB 180924 \citep[][]{2019Sci...365..565B}, FRB 190523 \citep[][]{ 2019Natur.572..352R} FRB 180916 \citep[][]{2020Natur.577..190M} , FRB20200120 \citep[][]{2021ApJ...910L..18B, 2021arXiv210511445K}, FRB 190520B \citep[][]{2022Natur.606..873N}. The other method uses the density profile and baryon fraction of host galaxy and halo to get the theoretical value of $\rm{DM_{host}}$, e.g. FRB 20181030A \citep[][]{2021ApJ...919L..24B}, FRB 190608 \citep[][]{2021ApJ...922..173C}. For the former method, modeling relations for $\rm{DM_{MW,ISM}, \ DM_{MW,Halo}}$ and the $\rm{DM_{IGM}-z}$ are applied to estimate $\rm{DM_{host}}$, while for the latter, various assumptions of electron distribution in host galaxies have been adopted. However, the uncertainty in these models could be significant for known localised events. For instance, $\rm{DM_{host}}$ of FRB20200120 could be smaller than $5 \rm{pc \, cm^{-3}}$, if $\rm{DM_{MW, Halo}} \sim 50-80 \, \rm{pc \, cm^{-3}}$ \citep[][]{2019MNRAS.485..648P}, instead of the value 30-40 $\rm{pc \, cm^{-3}}$ adopted in \cite{2021ApJ...910L..18B} and \cite{2021arXiv210511445K}.
 
In addition, some recent works report the estimation of the mean or median values of $\rm{DM_{host}}$ via statistical models of observed FRBs. \cite{2020Natur.581..391M} employs the dispersion measure of 7 localized FRB events to measure the baryon content of the universe. They assume that $\rm{DM_{host}}$ is following a log-normal distribution, and find a mildly favoured median value of $\sim 100\, \rm{pc \, cm^{-3}}$. Very recently, \cite{2022MNRAS.510L..18J} use the modeling of FRB observations to investigate the population of FRB, based on the events detected by ASKAP and Parkes, and find a best-fitting of the median $\rm{DM_{host}}$ is 130 $\rm{pc \, cm^{-3}}$. A median value of $\rm{DM_{host}} \gtrsim 100 \rm{pc \, cm^{-3}}$, favoured by \cite{2020Natur.581..391M} and \cite{2022MNRAS.510L..18J}, is more consistent with the result of the PSFR model based on the TNG galaxy samples. Nevertheless, we should be aware that $\rm{DM_{host}}$ generally increases with the stellar masses of hosts when $M_* < 10^{11} M_{\odot}$. The distribution and median value of $\rm{DM_{host}}$ in different stellar mass bins are desired to provide a more reliable constraint on the population of FRB.
 
Our work suggests that $\rm{DM_{host}}$ given by the PSFR and PStar population models can cover a broader range than that of localized events. Yet, the number of localized events that their $\rm{DM_{host}}$ are available is too small. Therefore, despite that the distribution of $\rm{DM_{host}}$ shows a notable difference between the two population models, current observations are insufficient to constrain the origin of FRB via the distribution of $\rm{DM_{host}}$. Moreover, giving the uncertainty in the current estimation on the $\rm{DM_{host}}$ of observed FRB events, it is practically challenge to obtain a reliable distribution of $\rm{DM_{host}}$ at the low DM end $\rm{DM_{host}} \lesssim 10 \rm{pc \, cm^{-3}}$ for observed FRBs. For instance, there is considerable scatter in the estimated $\rm{DM_{IGM}}$-z relation, which can be $\sim 30, 50, 100 \rm{pc\,cm^{-3}}$ at $z=0.2, 0.5, 1.0$ respectively (\citealt{2014ApJ...780L..33M, 2018ApJ...865..147Z, 2019MNRAS.484.1637J, 2021ApJ...906...95Z}). Consequently, if the $\rm{DM_{host}}$ of a localised events is derived by subtracting the foreground contribution, it will bear an uncertainty of order $\sim 30 \rm{pc\,cm^{-3}}$ at $z=0.2$. Therefore, it would be a difficult task to verify the difference between two population models in the distribution of $\rm{DM_{host}}$ at the low DM end. Alternatively, the cumulative probability at $\gtrsim 30 \rm{pc\,cm^{-3}}$ and the median value of $\rm{DM_{host}}$ would be more feasible indicators in the near future. 

\begin{figure}
    \includegraphics[width=1.\columnwidth]{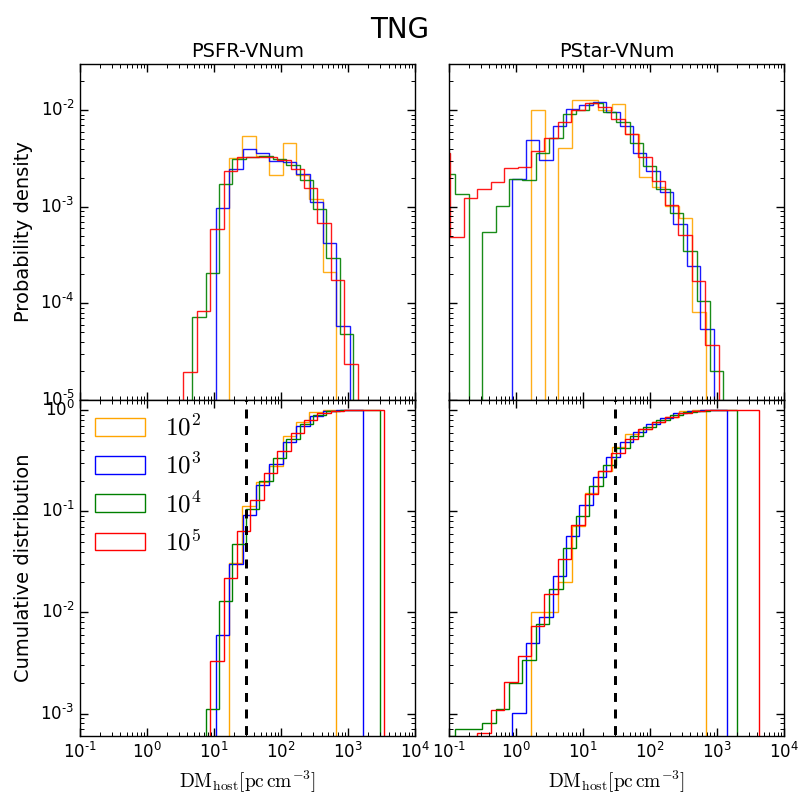}
    \caption{The probability density function (top) and cumulative distribution (bottom) of $\rm{DM_{host}}$ for a subgroup of mock FRB events in the TNG100-1 simulation at $z=0$. Blue, green, red histogram and solid lines are results of group sample size of $10^2, \, 10^3, \, 10^4, \, 10^5$ mock events respectively for the PSFR-VNum (left) and  PStar-VNum (right) models. The vertical dashed lines in the bottom row indicate $\rm{DM_{host}= 30 pc/cm^{3}}$.}
    \label{fig:compare totNumLOS}
\end{figure}

The number of localised events may increase to hundreds in the next few years, which will improve the statistical power. To provide a reference, we evaluate the number of events that could offer reliable statistics on the distribution and median value of $\rm{DM_{host}}$, and therefore can be able to constrain the different models considered in this work. We randomly select four subgroups of mock FRB events with the group sizes of $10^2, \, 10^3, \, 10^4, \, 10^5$ respectively from our whole FRB samples in the TNG simulation. Figure \ref{fig:compare totNumLOS} shows the probability and cumulative distribution function of $\rm{DM_{host}}$ of these four subgroups. For the case with $10^{3}$ events, the probability and cumulative distribution function at the low and high $\rm{DM_{host}}$ end is moderately suppressed with respect to the distribution of subgroups with $10^4$ and $10^5$ mock FRB events. This could partially explain why we have observed only a couple of FRB events with $\rm{DM_{host}}<10\,\rm{pc \, cm^{-3}}$ or $\rm{DM_{host}}>10^3\,\rm{pc \, cm^{-3}}$. On the other hand, the median value of $\rm{DM_{host}}$ is slightly changed when the number of mock events decreases from millions to 1000. To reduce the shot noise, We further repeat this procedure 20 times. We then measure the mean of the median $\rm{DM_{host}}$ in 20 subgroups for a certain group size, and its variance, which are listed in Table \ref{tab:mock FRBnum test}. The results suggest that a few  thousands of FRB events with known $\rm{DM_{host}}$ are sufficient to provide a statistical reliable description on the probability distribution of $\rm{DM_{host}}$.

\begin{table}
	\centering
	\caption{The mean of the median $\rm{DM_{host}}$ and its variance of 20 subgroups (group size N) of mock FRB events that are randomly selected from TNG100-1 simulation at $z=0$.  The subgroup sample size N changes from $10^2$ to $10^3, \, 10^4, \, 10^5$.}
	\label{tab:mock FRBnum test}
	\begin{tabular}{m{4em} m{5em} m{8em} m{4em} } 
		\hline
		model & mock FRB number  & mean of median $\rm{DM_{host}}$ [$\rm{pc\,cm^{-3}}$] & variance   \\
		\hline
		\multirow{4}{3em}{PSFR-VNum}  & $10^2$ & 172.72 & 258.25  \\
		                        & $10^3$ & 178.52 & 37.00  \\
		                        & $10^4$ & 179.27 & 6.46  \\
		                        & $10^5$ & 179.15 & 0.49  \\
		\hline
		\multirow{4}{3em}{PStar-VNum}  & $10^2$ & 64.99 & 185.86  \\
		                        & $10^3$ & 62.74 & 15.02  \\
		                        & $10^4$ & 63.39 & 0.78  \\
		                        & $10^5$ & 63.42 & 0.08  \\

		\hline
	\end{tabular}
\end{table}

Using the events detected by ASKAP and Parkes, \cite{2022MNRAS.510L..18J} suggests that the FRB population evolves with redshifts in a consistent way with the star formation rate, which supports a physical origin from young magnetars. However, as pointed out by \cite{2022MNRAS.510L..18J}, their results may not applicable to the total FRB population. For instance, \cite{2020MNRAS.498.3927H, 2022MNRAS.tmp..120H} demonstrated that rates of non-repeating FRB are likely to follow the cosmic stellar mass density evolution, indicating their progenitors would be old populations. Therefore, the distribution of $\rm{DM_{host}}$ of observed FRB in the future may be the mixture of the PSFR and PStar models. On the other hand, the offset of FRB events from galaxy center may also be used to ascertain the origin of FRB. Our study shows that there are moderate differences on the offset distances of mock events from host galaxies between the two different population models. However, our preliminary comparison indicates that the limited number of localized events to date makes it impossible to distinguish the two models at the present time. Besides, the stellar mass function of FRB hosts is also different between the two population models, which could be helpful to determine the origin of FRB. As Figure \ref{fig:galaxy mass function} shows, the event number weighted mass distribution of host galaxies for the Pstar model will have higher probability at high mass end, with respect to the PSFR model.

With more observed events in the future, incorporating statistics such as $\rm{DM_{host}}$, the offset from galaxy center and the host stellar mass function with existing measures will be of great benefit to ascertain the physical origin of FRBs. Meanwhile, the statistics of $\rm{DM_{host}}$ of localized FRB events could help constrain the sub-grid physics models that plays a key role in galaxy formation and evolution, because different sub-grid models can lead to significant discrepancy in the distribution of electron in the ISM and CGM of galaxies.

\section{Conclusions}
Based on galaxy samples in the IllustrisTNG and Illustris simulations, we make a comprehensive study on the dispersion measure of FRB contributed by the host galaxy and parent halo, $\rm{DM_{host}}$, between redshift z=0 and z=2. We have placed millions of mock FRB events in galaxies by two different FRB population models, one is assumed to trace the star formation rate, i.e. associated with young progenitor model, and the other is assumed to trace the stellar mass, i.e. associated with old progenitors. The former and latter are named as 'PSFR-VNum' and 'PStar-VNum' respectively. We investigate the distribution of $\rm{DM_{host}}$ to examine the dependency of $\rm{DM_{host}}$ on the stellar mass and morphology of host galaxies. We also study the redshift evolution of $\rm{DM_{host}}$. We find that:

There are significant differences in $\rm{DM_{host}}$ between the two FRB population models, and between galaxies in the IllustrisTNG and Illustris simulations as well. The distribution of $\rm{DM_{host}}$ deviates from a log-normal form for the both population models. For the model tracing stellar mass, it is due to an excess distribution at the low DM end ($\rm{ <10 \, pc\,cm^{-3} }$). Meanwhile, the median values of $\rm{DM_{host}}$ with the FRB population tracing SFR are 179.18 and 52.86 $\rm{pc \,cm^{-3}}$ for galaxies in the TNG100-1 and Illustris-1 simulations respectively, which are much larger than the corresponding values with the population tracing stellar mass. i.e., 63.41 and 31.32 $\rm{pc \,cm^{-3}}$. Moreover, for the FRB population model tracing stellar mass, $\rm{DM_{host}}$ of galaxies in the Illustris simulation shows a much higher and extended probability distribution at the low DM end than that in the TNG simulation.

At $z=0$, the medium outside the host galaxy but inside the host dark matter halo can be an important component of $\rm{DM_{host}}$, accounting $\sim 20-40\%$ for galaxies in the TNG. The contribution from the CGM in the host halo needs to be well accounted while estimating the redshift of FRB events with relatively smaller DM. $\rm{DM_{host}}$ increases with the stellar masses of host galaxy when $M_* < 10^{11} M_{\odot}$ for both the FRB population models, but shows fluctuations at the higher mass end. On the other hand, the distribution of $\rm{DM_{host}}$ looks similar between disk and non-disk galaxies, but the median value of $\rm{DM_{host}}$ for disk galaxies are higher by a fraction of $10-20\%$. For a Milky-Way like disk galaxy, the median value of the total DM caused by host galaxy and halo in our models, $\rm{DM_{host}}$, ranges from 65.45 to 212.93 $\rm{pc \,cm^{-3}}$ at $z=0$. As redshift increases, $\rm{DM_{host}}$ increases gradually in our study. The difference of $\rm{DM_{host}}$ between the two population models declines with increasing redshift, and vanishes at $z=2.0$.

The discrepancy on $\rm{DM_{host}}$ between the two population models at $z<2$ are largely caused by the difference of spatial distribution of mock FRB events within the host galaxies, i.e., offset from the center of host galaxies. FRB events produced by the PSFR-VNum model are more concentrated and locating in more central regions of galaxies, with respect to events produced by the PStar-VNum model. Comparing with the offset of 19 localized FRB, the PStar-VNum model is more favoured, which however needs more localized events to verify. The difference between the two population models narrows down with increasing redshifts due to the following reasons. The star formation rate in the Illustris and TNG simulation increases with redshifts and peaks at $z\sim 2$, which agrees with observations. Therefore, the star formation region in galaxies would highly coincident with the stellar mass in space at $z \sim 2$. The primary reason that leads to difference of $\rm{DM_{host}}$ between galaxy samples in the two simulations is that the electron number density profiles in galaxies are different. Within the radius of $r<10$ kpc where most mock FRB sources locate in, the electron number density, $n_e$, in galaxies from the TNG simulation is higher than that in the Illustris simulation, which probably results from the fact that the feedback model in the TNG simulation can effectively heat and hence ionize the gas within the central region of galaxies. 

Our estimations converge with respect to the number of L.O.S adopted for each FRB event, and to the segment length used for integrating along the ray in the DM calculation. Yet, we find that setting up a constant number of mock FRB events for each galaxy would significantly underestimate $\rm{DM_{host}}$, the median value may even be reduced by a factor of 1.4 - 2.4 if the host galaxies are within a stellar mass range of $10^{8-13}M_{\odot}$. Meanwhile, estimation based on a simulation with poor resolution will overestimate (underestimate) the $\rm{DM_{host}}$ of PSFR-VNum (PStar-VNum) models. Our study suggests that if the FRB events trace the star formation rate, a few events would coincident with others in spatial position, if the spatial resolution of observation tools is under 1 kpc. The corresponding probability is much smaller if FRB events trace the stellar mass. 

In the future, with more and more FRB events observed and localized, we could make more reliable restriction on FRB population models, by counting the distribution of $\rm{DM_{host}}$ in different ranges of stellar mass, and the distribution of the projected offset distance of FRB from galaxy center, as well as stellar mass function of hosts. In the near future, due to the uncertainties on the estimated $\rm{DM_{host}}$ of observed events, the cumulative distribution function at $\rm{DM_{host}} \sim 30-50\, \rm{pc \,cm^{-3}}$ and the median value of $\rm{DM_{host}}$ can serve as indicators to discriminate different population models. To this end, hundreds to a few thousands of events could provide reliable statistics. Meanwhile, with more reliable knowledge of $\rm{DM_{host}}$, we can constrain the baryonic feedback models in galaxies, which would improve our understanding of galaxy formation and evolution.

\section*{Acknowledgements}
We thank the anonymous referee for her/his useful comments
that improved the manuscript. This work is supported by the National Natural Science Foundation of China (NFSC) through grant 11733010. W.S.Z. is supported by NSFC grant 12173102.  Y.W. is supported by The Major Key Project of PCL. L.T. is supported by the NSFC grant No. 12003079. Analysis carried in this work was completed on the HPC facility of the School of Physics and Astronomy, Sun Yat-Sen University.

\section*{Data Availability}
The data underlying this article will be shared on reasonable request to the corresponding author.



\bibliographystyle{mnras}
\bibliography{host_dm} 




\appendix

\section{convergence tests}
\begin{table*}
	\centering
	\caption{The best fitting parameters of $\rm{DM_{host}}$ distribution and the median value for galaxies samples in the TNG simulations for all the tests mentioned in section 2.5.}
	\label{tab:fit param with DMtests}
	\begin{tabular}{m{14em} | m{7em}  m{4em}  m{4em} m{4em} m{5em} m{4em} m{3em}} 
		\hline
		tests & model & boundary & $\rm{\Delta L}$ = [kpc/h] & Num-L.O.S=  & median [$\rm{pc\,cm^{-3}}$] & $ e^{\mu}$[$\rm{pc\,cm^{-3}}$] &  $  \sigma  $ \\
		 \hline
		\multirow{2}{*}{`CNum' model} & PSFR-CNum & 1$\rm{R_{m200}}$ & 1.0 & 20  & 75.23 & 72.70  & 1.04 \\
		    & PStar-CNum & 1$\rm{R_{m200}}$ & 1.0 & 20  & 46.91 & 49.90  & 1.18 \\
		\cline{1-8}
		\multirow{2}{*}{halo boundary} & PSFR-VNum & 2$\rm{R_{m200}}$ & 1.0 & 20  & 186.63 & 172.06  & 0.87 \\
    		& PStar-VNum & 2$\rm{R_{m200}}$ & 1.0 & 20  & 74.13 & 75.13  & 1.14 \\
		\cline{1-8} 
		\multirow{6}{*}{$\rm{\Delta L}$} & \multirow{3}{*}{PSFR-VNum} & 1$\rm{R_{m200}}$ & 0.1 & 20  & 179.49 & 163.36  & 0.90 \\
	        & & 1$\rm{R_{m200}}$ & 0.5 & 20  & 178.34 & 162.18  & 0.90 \\
		    & & 1$\rm{R_{m200}}$ & 2.0 & 20  & 182.26 & 170.39 & 0.94 \\
		    & \multirow{3}{*}{PStar-VNum} & 1$\rm{R_{m200}}$ & 0.1 & 20  & 64.42 & 64.51  & 1.25 \\
	        & & 1$\rm{R_{m200}}$ & 0.5 & 20  & 64.02 & 64.10  & 1.25 \\
		    & & 1$\rm{R_{m200}}$ & 2.0 & 20  & 62.39 & 62.64  & 1.26 \\

		\cline{1-8} 
		\multirow{4}{*}{Number of L.O.S} & \multirow{2}{*}{PSFR-VNum} & 1$\rm{R_{m200}}$ & 1.0 & 10  & 178.80 & 163.10  & 0.91 \\
		    & & 1$\rm{R_{m200}}$ & 1.0 & 30  & 179.26 & 163.21  & 0.91 \\
		    & \multirow{2}{*}{PStar-VNum} & 1$\rm{R_{m200}}$ & 1.0 & 10  & 63.37 & 63.44  & 1.25  \\
		    & & 1$\rm{R_{m200}}$ & 1.0 & 30  & 63.36 & 63.42  & 1.25 \\
		    
        \cline{1-8}
		\multirow{2}{*}{TNG100-1 ($M_*=10^9-10^{11}\rm{M_{\odot}}$) } & PSFR-VNum & 1$\rm{R_{m200}}$ & 1.0 & 20  & 202.44 & 186.74  & 0.83 \\
		    & PStar-VNum & 1$\rm{R_{m200}}$ & 1.0 & 20  & 68.26 & 67.41  & 1.31 \\

        \cline{1-8}
		\multirow{2}{*}{TNG300-1 ($M_*=10^9-10^{11}\rm{M_{\odot}}$) } & PSFR-VNum & 1$\rm{R_{m200}}$ & 1.0 & 20  & 221.32 & 201.31  & 0.85 \\
		    & PStar-VNum & 1$\rm{R_{m200}}$ & 1.0 & 20  & 56.27 & 60.89  & 1.23 \\
		    
        \cline{1-8}
		\multirow{2}{*}{TNG100-3 ($M_*=10^9-10^{11}\rm{M_{\odot}}$) }& PSFR-VNum & 1$\rm{R_{m200}}$ & 1.0 & 20  & 285.16 & 245.41  & 0.82 \\
		    & PStar-VNum & 1$\rm{R_{m200}}$ & 1.0 & 20  & 55.56 & 63.22  & 1.13 \\

		\hline
	\end{tabular}
\end{table*}

To explore the convergence of our results, we have run some tests regarding factors such as the assumed number of mock FRB in each galaxy, the number of sightlines drawn for each mock FRB event, the total number of mock FRB events, the choice of segment length, as well as the simulation resolution and volume. The results of these tests are listed in Table ~\ref{tab:fit param with DMtests}.

\begin{figure*}
	\includegraphics[width=1.9\columnwidth]{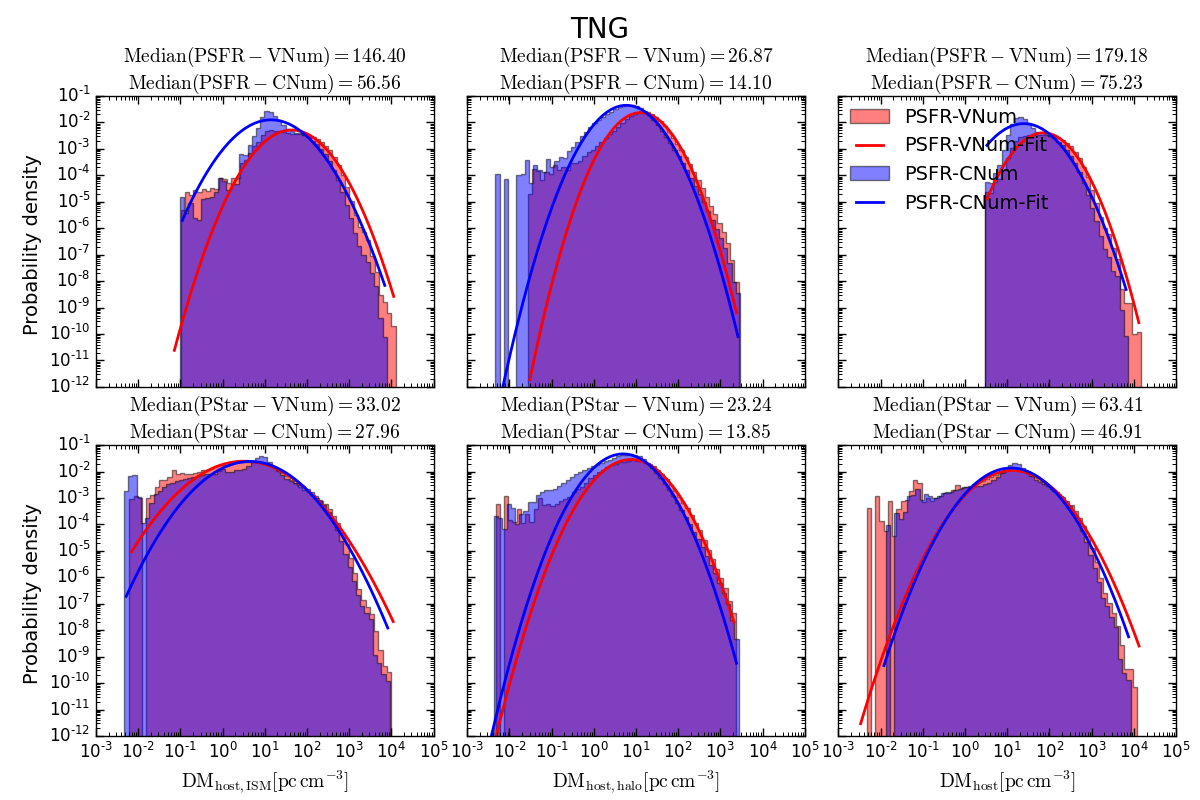}
    \caption{Top: The distribution and median of $\rm{DM_{host}}$ for the PSFR-VNum (red) and PSFR-CNum (blue) models with TNG100-1 samples. Bottom: same as top, but for the PStar-VNum and PStar-CNum models.}
    \label{fig:test of CNum model}
\end{figure*}

\begin{figure*}
	\includegraphics[width=1.6\columnwidth]{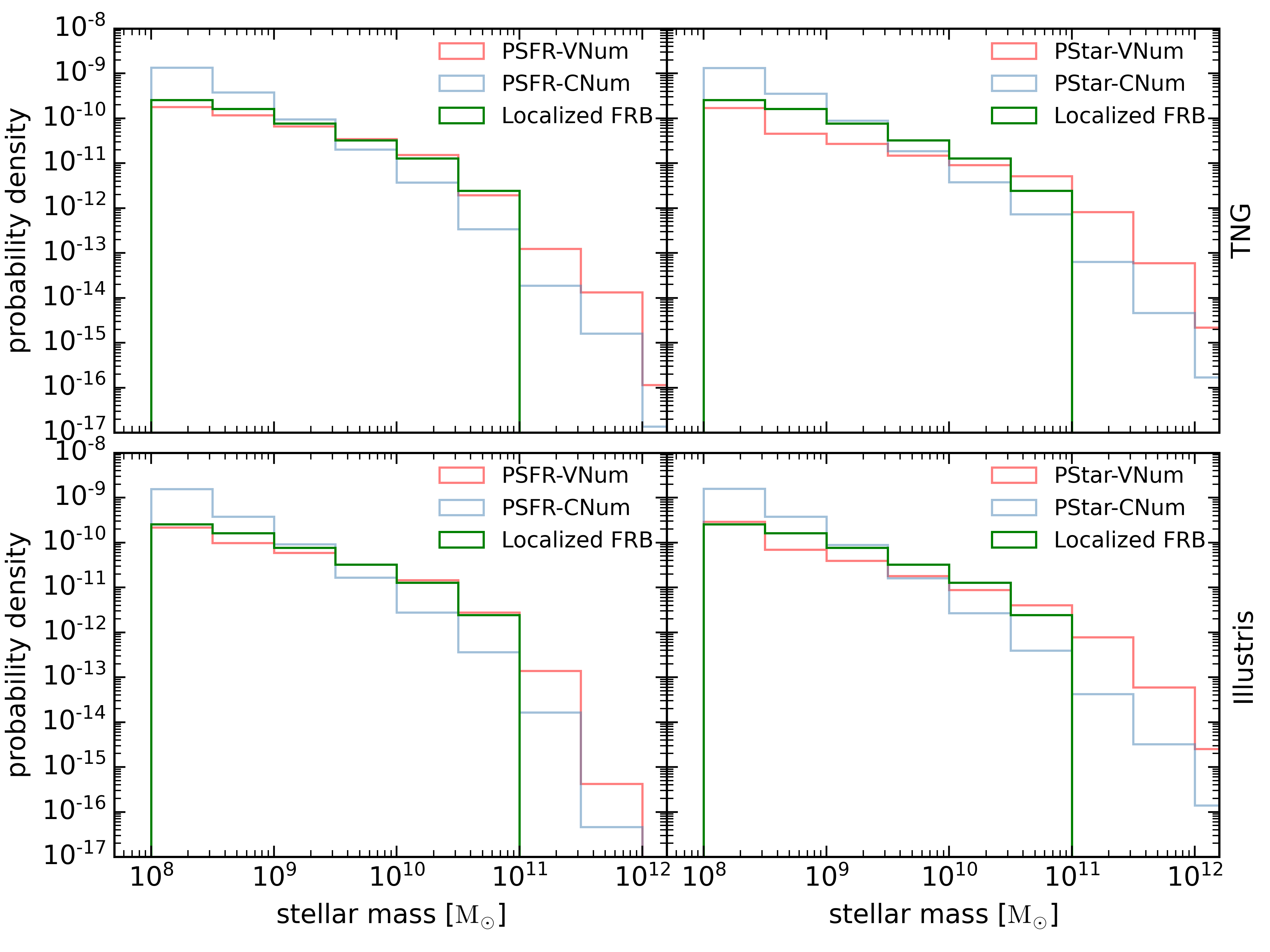}
    \caption{The red, blue, green lines indicate the stellar mass function of host galaxies that weighted by the number of FRB for Pxxx-VNum model, Pxxx-CNum model, and localized FRB samples respectively. Top(bottom) row is based on TNG100-1 (Illustris-1) simulation.}
    \label{fig:galaxy mass function}
\end{figure*}

\begin{figure*}
	\includegraphics[width=1.8\columnwidth]{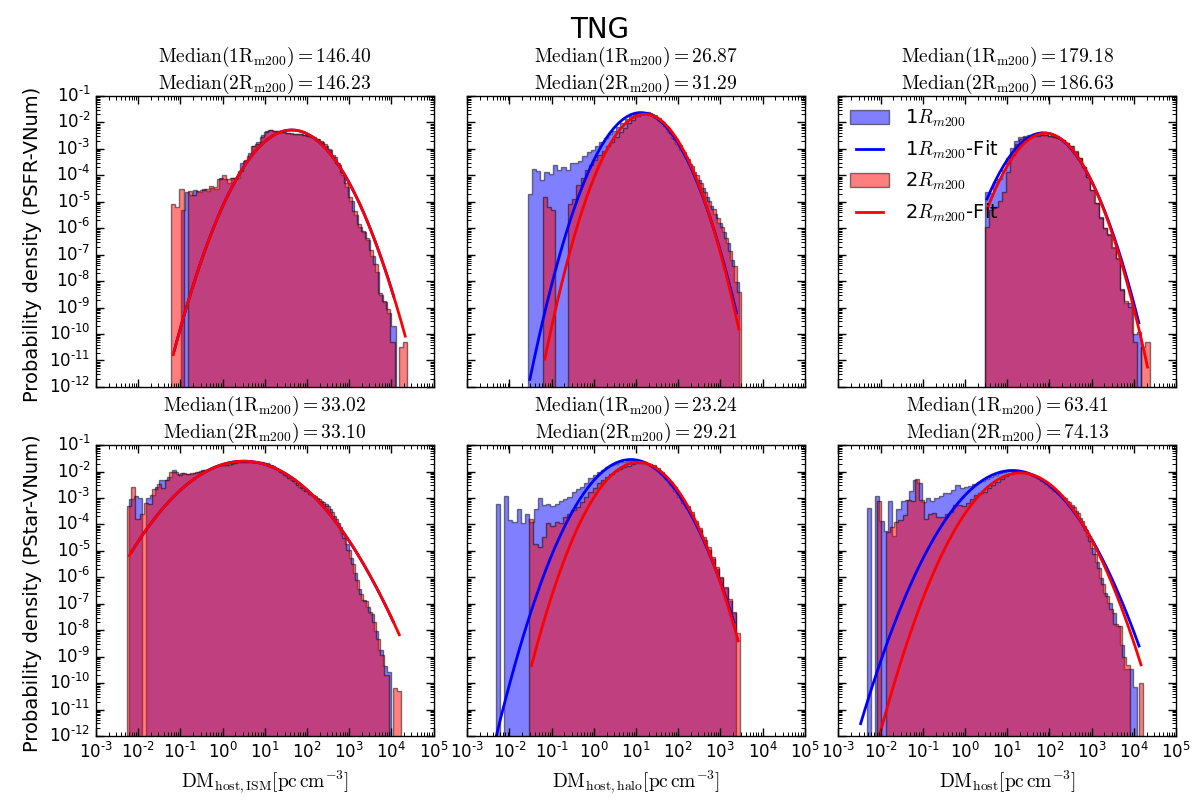}
    \caption{The distribution and median of $\rm{DM_{host}}$ for the PSFR-VNum (top) and PStar-VNum (bottom) models with TNG100-1 samples. Blue (red) indicates results with halo radius of $R_{m200}$ ( 2$R_{m200}$) .}
    \label{fig:test of 2RM200}
\end{figure*}

We would like firstly draw the readers' attention to the impact of the assumed FRB number in each galaxy. \cite{2020ApJ...900..170Z} adopt a constant number of 500 FRB events in each host galaxy. Our results presented in Section \ref{sec:DMs main results} are based on the assumption that the number of FRB events assigned to a given galaxy is related to either its stellar mass or SFR in the 'VNum' scenario. As introduced in Section \ref{sub:FRB population}, we have also considered the 'CNum' scenario in which a constant number of 12 FRBs is assigned to each galaxy. Figure \ref{fig:test of CNum model} shows the distribution of $\rm{DM_{host}}$ in the 'CNum' scenario and their median values. We also fit the distribution with the log-normal formula, and the fitting  parameters are listed in Table \ref{tab:fit param with DMtests}. For galaxies selected from both the two simulations, the value of $\rm{DM_{host}}$ in the 'CNum' scenario is significantly smaller than that in the 'VNum' scenario. More specifically, the 'PSFR-CNum' model predicts a median $\rm{DM_{host}}$ of 75.23 $\rm{pc \,cm^{-3}}$ for the TNG100-1 samples, against 179.18 $\rm{pc \,cm^{-3}}$ in the 'PSFR-VNum' model. The 'PStar-CNum' model results in a median $\rm{DM_{host}}$ of 46.91, $\rm{pc \,cm^{-3}}$ for the TNG100-1 samples, but the corresponding value in the 'PStar-VNum' model are 63.41 $\rm{pc \,cm^{-3}}$ respectively.

This is not out of expectation, as mock FRB are more likely hosted by massive galaxies in the 'VNum' scenario. With a constant number of FRB events for each galaxy, the FRB number weighted mass distribution of host galaxy would follow exactly the original galaxy stellar mass function in simulations, dominated by low mass galaxies. However, if the number of FRB is related with stellar masses or SFR model, the event number weighted mass distribution of host galaxies will lean toward the high mass end, as shown by Figure \ref{fig:galaxy mass function}. Despite that the number of localised FRB events is very limited, the stellar mass function of their hosts shows a better agreement with the 'VNum' scenario as indicated by Figure \ref{fig:galaxy mass function}. 

Figure ~\ref{fig:test of 2RM200} compares the distribution of $\rm{DM_{host}}$ with two different definitions of halo boundary, the virial radius $
\rm{R_{m200}}$ against $2
\rm{R_{m200}}$. Increasing the halo radius leads to the probability at the low DM end decrease moderately, and hence increases the median $\rm{DM_{host}}$. In addition, we find that there will be barely any difference on the distribution and median value of $\rm{DM_{host}}$ if we draw 30, instead of 20, sightlines for each FRB event. We further find this change have minor effect on the distribution of $\rm{DM_{host}}$ in disk galaxy samples (results are not shown for the sake of conciseness). Therefore, 20 L.O.S for each FRB is enough to obtain reliable statistical results.

Furthermore, we probe the impact of the choice of $\Delta L$ in the eqn. \ref{eqn:dm_diff} by using different values of $\Delta L= 0.1, \, 0.5, \, 2.0\, \rm{kpc}/h$, which result in minor variations on the distribution of $\rm{DM_{host}}$ and its median value. The corresponding values can be found in Table \ref{tab:fit param with DMtests}. Hence, our choice of $\Delta L=1 \, \rm{kpc}/h$ is reasonable.

\begin{figure*}
	\includegraphics[width=1.8\columnwidth]{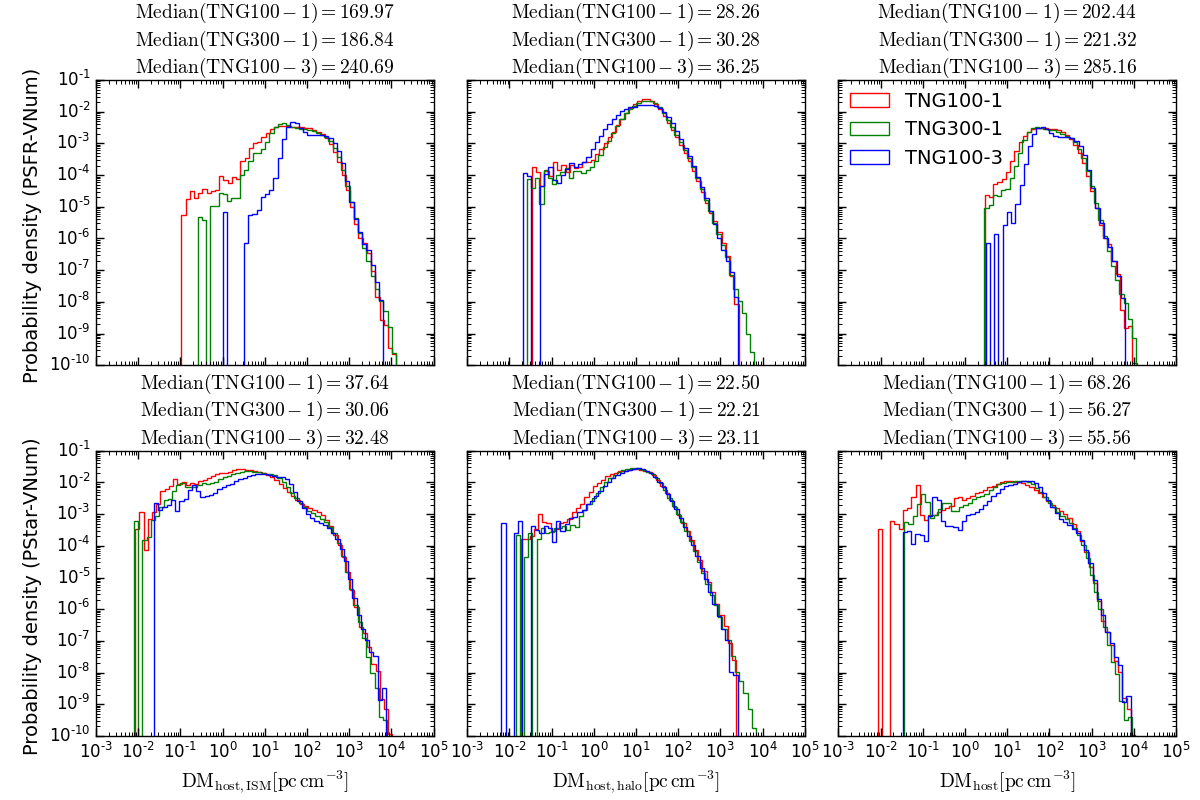}
    \caption{The distribution and median of $\rm{DM_{host}}$ for galaxies samples in the TNG100-1, TNG100-3 and TNG300-1 simulations.}
    \label{fig:test of TNG100-3 and TNG300-1}
\end{figure*}

\begin{figure*}
	\includegraphics[width=1.3\columnwidth]{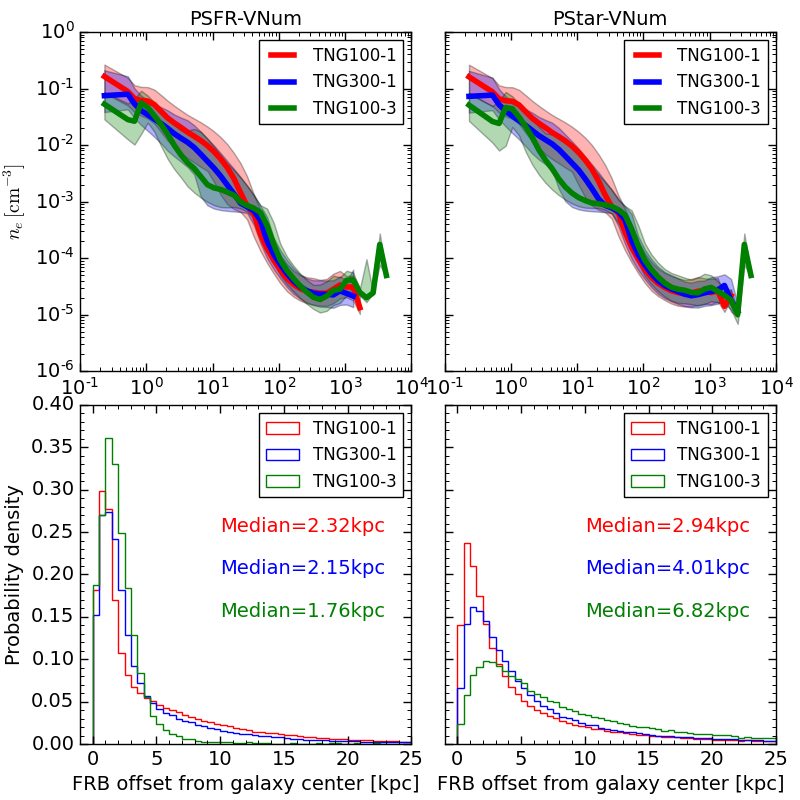}
    \caption{The $n_e$ profile (top), projected FRB distribution and its median distance (bottom) for galaxies samples with stellar mass range $10^9-10^{11}M_{\odot}$ in the TNG100-1, TNG100-3 and TNG300-1 simulations.}
    \label{fig:ne profile in TNG100-3 and TNG300-1}
\end{figure*}

Finally, we have investigated how the simulation resolution and volume will influence our results. We have applied our calculation to galaxies samples in the TNG300-1 and TNG100-3 simulations. TNG300-1 has box size of $(300)^3 \, \rm{Mpc}$, but has a mass resolution lower than TNG100-1 by 8 times. TNG100-3 has the same volume as TNG100-1, but the mass resolution is lower by 64 times. To make a fair comparison, we select galaxies in the same stellar mass range, i.e., $10^9-10^{11} \rm{M_{\odot}}$, from these three simulations. Figure \ref{fig:test of TNG100-3 and TNG300-1} shows the distribution of $\rm{DM_{host}}$. For the PSFR-VNum model, the median $\rm{DM_{host}}$ of samples in simulations with higher resolution is lower, and decreases from 285.16 $\rm{pc/cm^3}$ in TNG100-3 to 202.44 $\rm{pc/cm^3}$. The reverse trend is observed for the PStar-VNum model, but the difference narrows to $\sim 20\%$ between TNG100-3 and TNG100-1. These trends are caused by the changes on the distribution of electron density and mock events. As shown by the top row in Figure \ref{fig:ne profile in TNG100-3 and TNG300-1}, the electron density in the region with $r\lesssim 30 \rm{kpc}/h$ decreases from TNG100-1 to TNG300-1, and then to TNG100-3. Meanwhile, the SFR region has a more extended distribution in a simulation with higher resolution. However, the stellar component is more concentrate when the resolution increases. On the other hand, the results of TNG300-1 fall between that of TNG100-1 and TNG100-3. As the difference between TNG300-1 and TNG100-1 is partly caused by the resolution, the impact of simulation volume should be limited.

\section{Some extra material}
\begin{table*}
	\centering
	\caption{Results of fitting with different fitting range.The best fitting parameters and median value of $\rm{DM_{host}}$ and its two components $\rm{DM_{host, ISM}}$ and $\rm{DM_{host, halo}}$, derived from galaxies samples in the TNG100-1 simulations by different models.}
	\label{tab:fit range test}
	\begin{tabular}{m{7em} m{6em} m{5em} m{5em} m{5em} m{4em} m{8em}}  
		\hline
    		simulation & model & component & median [$\rm{pc\,cm^{-3}}$] & $ e^{\mu}$[$\rm{pc\,cm^{-3}}$] &  $ \sigma$ & mean  p-value(1000 mock events)  \\
		\hline
		\multirow{6}{*}{TNG(DM>1)} & \multirow{3}{*}{PSFR-VNum} & $\rm{DM_{host,ISM}}$  & 146.41 & 126.98  & 1.03 & 0 \\
		      &   & $\rm{DM_{host,halo}}$  & 26.89 & 28.27  & 0.88 & 0.003 \\
		      &   & $\rm{DM_{host}}$  & 179.18 & 163.23 & 0.91 & 0 \\
		                        
		      & \multirow{3}{*}{PStar-VNum} & $\rm{DM_{host,ISM}}$  & 33.92 & 36.45 & 1.47 & 0\\
		      &   & $\rm{DM_{host,halo}}$ & 23.32 & 23.38 & 1.04 & 0.12 \\
		      &   & $\rm{DM_{host}}$ & 63.62 & 64.17 & 1.23 & 0.008 \\
		\hline

		\multirow{6}{*}{TNG(DM>10)} & \multirow{3}{*}{PSFR-VNum} & $\rm{DM_{host,ISM}}$  & 148.74 & 131.61  & 0.99 & 0  \\
		      &   & $\rm{DM_{host,halo}}$  & 29.87 & 33.49  & 0.76 & 0 \\
		      &   & $\rm{DM_{host}}$  & 179.34 & 163.61 & 0.90 & 0 \\
		                        
		      & \multirow{3}{*}{PStar-VNum} & $\rm{DM_{host,ISM}}$  & 54.32 & 61.58 & 1.15 & 0 \\
		      &   & $\rm{DM_{host,halo}}$ & 29.61 & 33.54 & 0.80 & 0 \\
		      &   & $\rm{DM_{host}}$ & 70.88 & 75.68 & 1.08 & 0 \\
		\hline

		\multirow{6}{*}{TNG(DM>30)} & \multirow{3}{*}{PSFR-VNum} & $\rm{DM_{host,ISM}}$  & 167.37 & 161.43  & 0.80 & 0 \\
		      &   & $\rm{DM_{host,halo}}$  & 52.00 & 60.86  & 0.60 & 0  \\
		      &   & $\rm{DM_{host}}$  & 189.26 & 180.85 & 0.80 & 0 \\
		                        
		      & \multirow{3}{*}{PStar-VNum} & $\rm{DM_{host,ISM}}$  & 109.36 & 115.44 & 0.87 & 0 \\
		      &   & $\rm{DM_{host,halo}}$ & 53.81 & 63.41 & 0.62 & 0 \\
		      &   & $\rm{DM_{host}}$ & 104.08 & 115.00 & 0.86 & 0 \\
		\hline
		\hline
	\end{tabular}
\end{table*}

\begin{figure*}
	\includegraphics[width=\columnwidth]{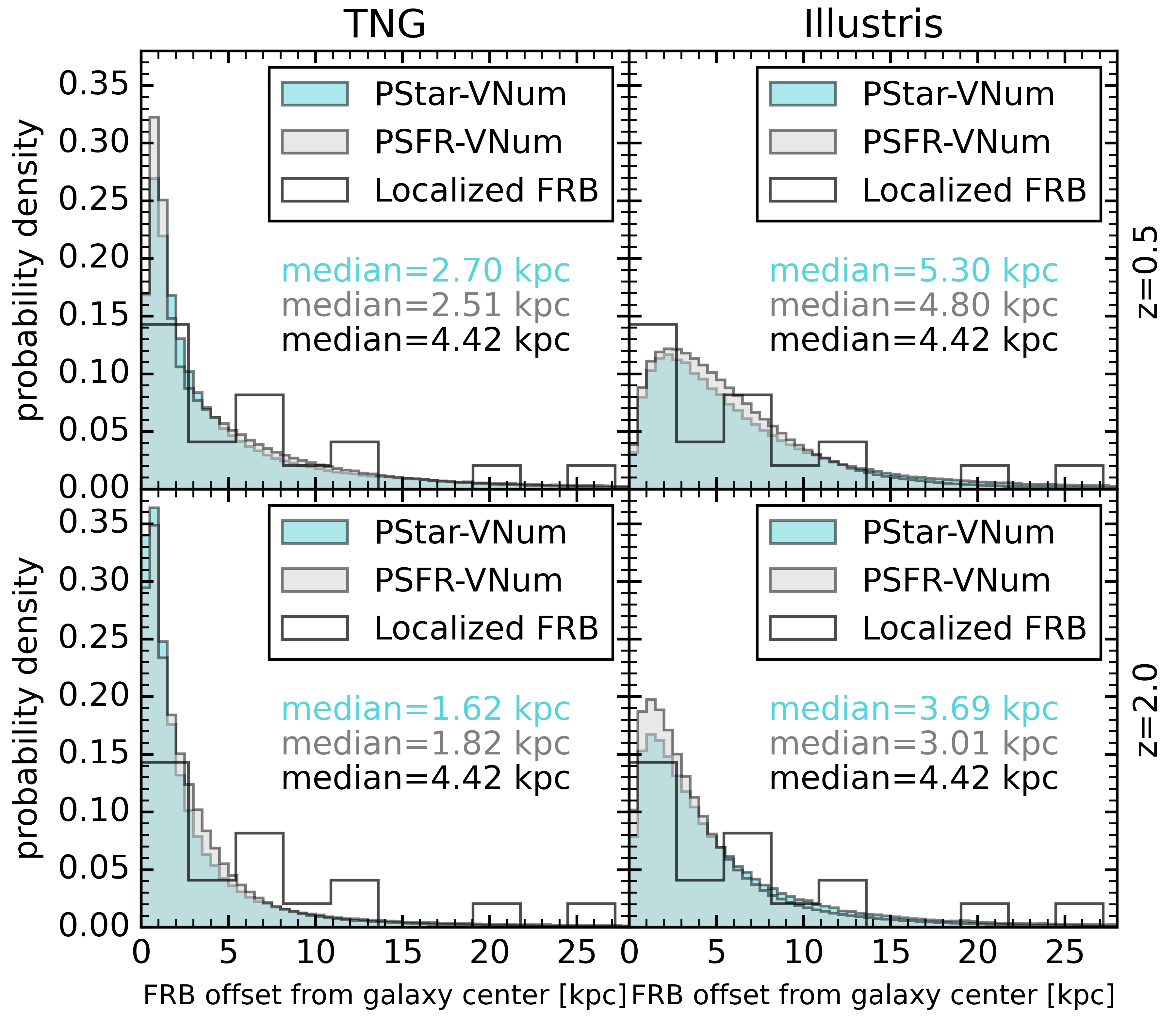}
    \caption{The distribution of the projected offset distance of FRB events from host galaxy center of two models and localized FRBs. Left (Right): Results based on galaxies samples in the TNG100-1 (Illustris-1) simulation. Top (Bottom): the results at z=0.5 (2.0). The blue, red histograms and texts show the distribution and median value of offset distance with 'PStar-VNum' and 'PSFR-VNum' model, respectively. The white histograms and black text are the distribution and median value of 19 observed FRBs whose host galaxies have been localized.}
    \label{fig:distance distribution z0.5}
\end{figure*}

\begin{figure*}
	\includegraphics[width=1.6\columnwidth]{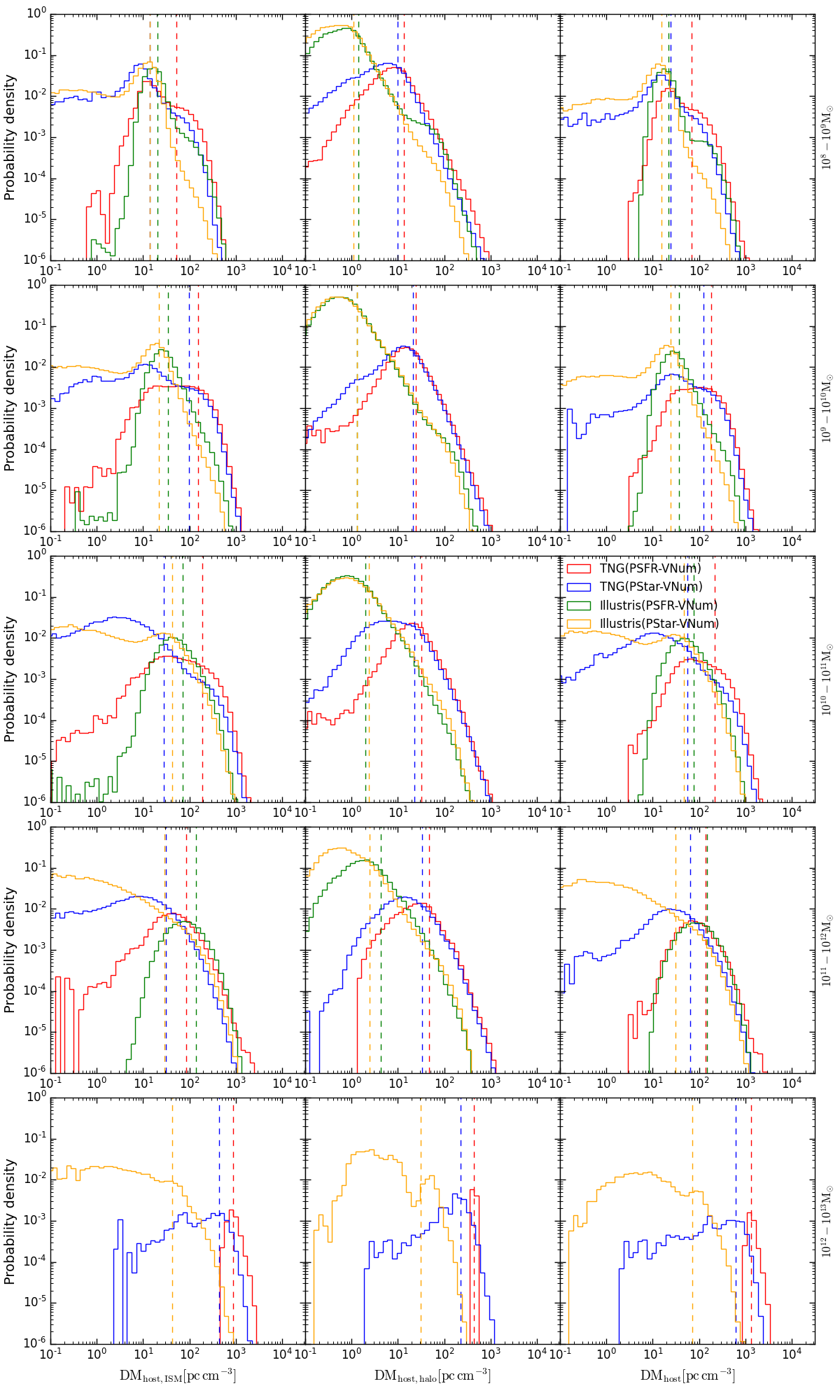}
    \caption{
    The probability distribution of $\rm{DM_{host}}$ in different host stellar mass bins (labeled on the right axis) at $z=0.0$ . The left, middle and right column are DM distribution of 'ISM', 'Halo' and 'Host' components respectively. Red, blue, green, orange histogram (vertical dashed lines) are distribution (median DM) of 'PSFR-VNum' and 'PStar-VNum'  model in TNG100-1, 'PSFR-VNum' and model in Illustris-1 respectively.}
    \label{fig:tng massbin pdf of two models with 5R50}
\end{figure*}



\bsp	
\label{lastpage}
\end{document}